\documentclass{article}
\usepackage{graphicx} 
\usepackage[a4paper,margin=1in]{geometry}

\usepackage{graphicx}
\usepackage{subcaption}
\usepackage{float}
\usepackage{placeins}

\usepackage{amsmath}
\usepackage{amssymb}
\usepackage{mathtools}
\usepackage{bm}

\usepackage{booktabs}
\usepackage{multirow}
\usepackage{array}
\usepackage{tabularx}

\usepackage{siunitx}

\usepackage[numbers,sort&compress]{natbib}
\usepackage[colorlinks=true,
linkcolor=blue,
citecolor=blue,
urlcolor=blue]{hyperref}
\usepackage[nameinlink,capitalise]{cleveref}

\usepackage{caption}
\usepackage{indentfirst}
\usepackage{lineno}
\usepackage{authblk}

\title{Scale-aware composite forcing enhances forward energy transfer in lattice Boltzmann simulations of forced homogeneous isotropic turbulence
}

\author[1]{Hongxuan Zhang}
\author[1]{Fanxi Gong}
\author[1]{Yating Chen}
\author[2]{Runchao Guan}
\affil[1]{School of Atmospheric Sciences, Sun Yat-sen University, Zhuhai, 51900, Guangdong, China.}
\affil[2]{School of Mechatronics Engineering, Harbin Institute of Technology, Harbin 150001, China}
\affil[1]{\textit{Email: zhanghx239@mail2.sysu.edu.cn}}
\date{June 2026}

\begin{document}

\maketitle

\begin{abstract}

In simulations of forced homogeneous isotropic turbulence (FHIT), large-scale forcing can maintain a statistically stationary state. However, the intermediate-scale receives energy mainly through nonlinear transfer, limiting targeted control over energy levels within a prescribed band. This work proposes a scale-aware composite forcing strategy for lattice Boltzmann method. Composite filtering extracts the velocity field for intermediate-scale feedback, and the feedback force is constructed along its local direction. The magnitude of the cross-power spectrum between the intermediate-scale feedback and the filtered velocity is concentrated primarily in the target band, with a time-averaged band fraction of \(72.26\%\)-\(74.56\%\). Intermediate-scale feedback increases energy in both the target and high-wavenumber bands, with a larger relative increase in the latter. Compared with the baseline without intermediate-scale feedback, the time-averaged relative increases in the target-band energy fraction are \(6.29\%\), \(11.65\%\), \(18.99\%\), and \(19.89\%\) for \(\alpha=1\%\), \(3\%\), \(5\%\), and \(7\%\), respectively. For these four feedback coefficients \(\alpha\), the mean nonlinear energy flux across \(K=12\) reaches \(1.18\), \(1.54\), \(2.32\), and \(3.06\) times the baseline value, respectively. The relative increases are greater than those across \(K=4\) in all four cases, indicating scale-dependent enhancement of net forward nonlinear energy transfer. As feedback coefficient \(\alpha\) increases from 0\% to 7\%, the mean Taylor-microscale Reynolds number \(Re_\lambda\) rises from \(73.94\) to \(97.86\), enstrophy increases by \(181.98\%\) relative to the baseline and the skewness of the longitudinal velocity gradient shifts toward zero, while the standardized velocity distributions and normalized \(Q'\)--\(R'\) joint distributions largely retain their overall shapes. On the current grid, the mean \(k_{\max}\eta\) values for the \(5\%\) and \(7\%\) feedback cases are \(1.01\) and \(0.90\), respectively, indicating that the former approaches the resolution limit, whereas the latter is at risk of under-resolution. In a comparison at approximately matched mean dissipation rates, the feedback coefficient \(5\%\) case yields a mean target-band energy fraction of \(32.36\%\), compared with \(27.55\%\) for the enhanced large-scale forcing case. After the initial transient, its net forward nonlinear energy flux across \(K=12\) is generally \(2.17\)--\(2.39\) times that of the comparison case. This composite forcing strategy provides a viable approach to targeted control of energy distribution and energy transfer across scales in FHIT, with feedback coefficient \(\alpha\) chosen to balance target-band energy gains against small-scale resolution constraints. Future work could incorporate grid refinement to further assess the applicability of this strategy at different Reynolds numbers and in different target bands, and explore adaptive adjustment of feedback strength based on energy response and resolution constraints.

\end{abstract}

\noindent\textbf{Keywords:} Forced homogeneous isotropic turbulence; Lattice Boltzmann method; Composite forcing; Intermediate-scale feedback; Nonlinear energy transfer.

\section{Introduction}

Multiscale energy transfer is a central problem in turbulence research~\cite{kolmogorov1991local,pope2001turbulent}. Turbulent kinetic energy is typically injected or generated on large scales and transferred to smaller scales through nonlinear interactions before being dissipated by viscosity~\cite{kolmogorov1995turbulence}. In geophysical flows, buoyancy, rotation, stable stratification, topography, and boundary conditions modulate this transfer~\cite{wyngaard2010turbulence,vallis2017atmospheric}. These influences complicate efforts to isolate small-scale statistics and the mechanisms that govern their development. Homogeneous isotropic turbulence (HIT) excludes mean shear and wall effects, providing an idealized framework for studying turbulence~\cite{batchelor1953theory}. In freely decaying HIT, the turbulent kinetic energy decreases over time through viscous dissipation. Numerical simulations therefore commonly employ external forcing to compensate for this loss and maintain statistically stationary turbulence, producing forced homogeneous isotropic turbulence (FHIT)~\cite{eswaran1988examination,alvelius1999random}. FHIT currently serves as a benchmark for evaluating turbulence simulations, investigating energy cascade dynamics, and assessing forcing strategies~\cite{rosales2005linear}.

In FHIT simulations, the forcing strategy determines where energy is injected and influences statistically stationary flow statistics and energy transfer across scales~\cite{bassenne2016constant}. Traditional low-wavenumber forcing supplies energy at large scales to sustain turbulence, while nonlinear interactions transfer this energy to intermediate and small scales~\cite{doering2005low,mccomb1974local,ten2006application}. This approach offers limited direct control over energy input within a prescribed intermediate-scale band. Intermediate scales connect large-scale energy input with small-scale dissipation and are therefore important for studying and modulating interscale energy transfer~\cite{alexakis2018cascades,eyink2009localness,johnson2020energy}. Linear Physical-space forcing provides a straightforward feedback formulation based on the instantaneous velocity field. However, feedback applied to the full velocity field without scale decomposition does not explicitly select a target band~\cite{domaradzki1990local,lallemand2000theory,kuczaj2006nonlocal}. Combining scale extraction with feedback forcing may enable the targeted control of intermediate-scale motions. This scale-selective feedback could provide a way to promote energy transfer~\cite{waleffe1992nature,carroll2013proposed,aluie2009localness}.

The lattice Boltzmann method (LBM) uses a local collision--streaming scheme with a simple algorithmic structure and is well suited to parallel computation~\cite{timm2016lattice}. These features facilitate large-scale three-dimensional turbulence simulations in regular periodic domains. Previous studies have reported good agreement with the spectral methods for HIT energy spectra and key statistics under low-Mach-number conditions and with adequate spatial resolution~\cite{latt2021cross}. LBM can also apply and combine different forms of forcing strategies and comparisons of flow responses ~\cite{karlin2003entropic,moin1998direct}. Together, these capabilities make LBM suitable for FHIT simulations, the collection of stationary turbulence statistics, and investigations of energy transfer across scales~\cite{yeung2018effects,mattila2017high}.

This work proposes a combined large- and intermediate-scale forcing method for FHIT simulations within the LBM framework. The method keeps low-wavenumber forcing to maintain turbulence and uses composite physical-space filtering to extract target intermediate-scale motions. The amplitude-limited feedback is then applied along the direction of the filtered velocity. We assess the scale selectivity of the constructed coupling using the cross-power spectrum between the feedback force and the filtered velocity. Energy spectra and band-integrated energies characterize the resulting changes in the scale-wise energy distribution. Cumulative nonlinear energy fluxes are then examined to determine whether these changes accompany stronger net forward nonlinear energy transfer. We further characterize the flow response using Taylor-microscale Reynolds number \(Re_\lambda\), enstrophy, skewness, standardized velocity PDFs, and normalized \(Q'\)--\(R'\) joint distributions. The small-scale resolution indicator \(k_{\max}\eta\) is used to assess the constraints on the choice of feedback strength under current computational conditions. Finally, a large-scale-only control with approximately matched mean dissipation tests whether the target-band energy and nonlinear flux enhancements merely result from increased total energy input.

\section{Numerical method and composite forcing model}
This work simulates the FHIT using a weakly compressible D3Q19 lattice Boltzmann model with the Bhatnagar--Gross--Krook (BGK) single-relaxation-time collision operator~\cite{qian1992lattice,bhatnagar1954model}. The standard D3Q19-BGK solver is retained, with modifications confined to the forcing term. Intermediate-scale feedback is introduced, operating alongside the traditional low-wavenumber, large-scale forcing. We first present the LBM evolution equation and the recovery of macroscopic variables. Then describe the large-scale forcing, the intermediate-scale velocity extraction, and the combined forcing formulation.

\subsection{Single-relaxation-time lattice Boltzmann model}
\label{subsec:srt_lbm}
Within the D3Q19-BGK framework, the evolution of the distribution function $f_i(\mathbf{x},t)$ is governed by\cite{bhatnagar1954model,qian1992lattice,chen1998lattice,mcnamara2019use,higuera1989boltzmann}:
\begin{equation}
f_i\left(\mathbf{x}+\boldsymbol{e}_i\delta_t,t+\delta_t\right)
=f_i(\mathbf{x},t)-\frac{1}{\tau}\left[f_i(\mathbf{x},t)
-f_i^{\mathrm{eq}}(\mathbf{x},t)\right]+F_i(\mathbf{x},t),
\qquad i=0,1,\ldots,18.
\label{eq:lbe_bgk}
\end{equation}
Here, $\boldsymbol{e}_i$ is the discrete lattice velocity and $\delta_t$ is the time step. The dimensionless relaxation time is denoted by $\tau$, and $f_i^{\mathrm{eq}}$ is the local equilibrium distribution function. The discrete forcing term associated with the lattice direction $i$ is denoted by $F_i$.

For the D3Q19 model, the discrete velocity set is defined as:
\begin{equation}
\mathbf{e}_i=\begin{cases}(0,0,0), & i=0, \\[3pt]
(\pm 1,0,0),\ (0,\pm 1,0),\ (0,0,\pm 1), & i=1,\ldots,6, \\[3pt]
(\pm 1,\pm 1,0),\ (\pm 1,0,\pm 1),\ (0,\pm 1,\pm 1), & i=7,\ldots,18.
\end{cases}
\label{eq:d3q19_velocity}
\end{equation}
The corresponding weighting coefficients are
\begin{equation}
w_i=\begin{cases}\dfrac{1}{3}, & i=0, \\[6pt]\dfrac{1}{18}, & i=1,\ldots,6, \\[6pt]
\dfrac{1}{36}, & i=7,\ldots,18.\end{cases}\label{eq:d3q19_weights}
\end{equation}
All simulations use lattice units with $\delta_x=\delta_t=1$. The lattice speed is therefore $c=\delta_x/\delta_t=1$, and the lattice sound speed satisfies $c_s^2=1/3$.

The local equilibrium distribution function is obtained from a second-order expansion of the Maxwell--Boltzmann distribution~\cite{qian1992lattice,he1997theory}:
\begin{equation}
f_i^{\mathrm{eq}}(\rho,\mathbf{u})=\rho w_i\left[
1+\frac{\boldsymbol{e}_i\cdot\mathbf{u}}{c_s^2}
+\frac{\left(\boldsymbol{e}_i\cdot\mathbf{u}\right)^2}{2c_s^4}
-\frac{\mathbf{u}\cdot\mathbf{u}}{2c_s^2}\right],
\label{eq:equilibrium_distribution}
\end{equation}
where $\rho$ and $\mathbf{u}$ are the macroscopic density and velocity, respectively.
The macroscopic density is recovered from the zeroth-order moment of the distribution functions, which is shown as:
\begin{equation}
\rho(\mathbf{x},t) = \sum_{i=0}^{18} f_i(\mathbf{x},t).
\label{eq:macroscopic_density}
\end{equation}
The macroscopic momentum is obtained from the first-order moment,
\begin{equation}
\rho\mathbf{u}(\mathbf{x},t)=\sum_{i=0}^{18}
\boldsymbol{e}_i f_i(\mathbf{x},t).
\label{eq:macroscopic_momentum}
\end{equation}
For the weakly compressible isothermal model, pressure and density are related by the equation of state $p=c_s^2\rho$.
Viscosity $\nu$ is related to relaxation time $\tau$ by:
\begin{equation}
\nu = \frac{1}{3}\left(\tau-0.5\right).
\label{eq:kinematic_viscosity_lu}
\end{equation}
The macroscopic forcing acceleration $\mathbf{F}$ enters the evolution of a distribution function through the discrete forcing term~\cite{succi2001lattice,guo2002discrete}:
\begin{equation}
F_i = 3\rho w_i\left(\boldsymbol{e}_i\cdot\mathbf{F}\right).
\label{eq:discrete_forcing}
\end{equation}
In the composite forcing framework, the total macroscopic forcing is decomposed as follows:
\begin{equation}\mathbf{F} = \mathbf{F}_{\mathrm{L}}+\mathbf{F}_{\mathrm{mid}}.
\label{eq:total_forcing}
\end{equation}
Here, $\mathbf{F}_{\mathrm{L}}$ is the large-scale that maintains statistically stationary turbulence, and $\mathbf{F}_{\mathrm{mid}}$ is the scale-selective intermediate-scale feedback. The formulations of both contributions are detailed in the following subsections.

\subsection{Large-scale forcing for statistically stationary FHIT}
To maintain statistically stationary FHIT, this work proposes large-scale forcing at low wavenumbers. It supplies the initial energy input and is denoted by $\mathbf{F}_{\mathrm{L}}=(F_x,F_y,F_z)$. Following the forcing formulation of Kareem et al.~\cite{kareem2022simulations}, its three components are defined as:
\begin{equation}
\mathbf{F}_{\mathrm{L}} \equiv
\left\{\begin{aligned}
F_x &=A \sum_{k_i=-k_{\max}}^{k_{\max}}\left[\left(\frac{k_2k_3}{k^2}\right)\sin\left(\frac{2\pi}{N}(k_1x+k_2y+k_3z)+\phi_{\mathbf{k}}\right)\right], \\[4pt]
F_y &=-2A \sum_{k_i=-k_{\max}}^{k_{\max}}\left[\left(\frac{k_1k_3}{k^2}\right)\sin\left(\frac{2\pi}{N}(k_1x+k_2y+k_3z)+\phi_{\mathbf{k}}\right)
\right], \\[4pt]
F_z &=A \sum_{k_i=-k_{\max}}^{k_{\max}}\left[\left(\frac{k_1k_2}{k^2}\right)\sin\left(\frac{2\pi}{N}(k_1x+k_2y+k_3z)+\phi_{\mathbf{k}}\right)\right],
\end{aligned}
\right.
\label{eq:large_scale_forcing}
\end{equation}
where $A$ is the large-scale forcing amplitude and $N$ is the number of grids  in each spatial direction. The forcing acts on modes satisfying $0<k\leq k_{\max}$, where $k=\sqrt{k_1^2+k_2^2+k_3^2}$ and $k_{\max}=4$ in the present simulations. A random phase $\phi_{\mathbf{k}}$ is assigned to each forced mode. The zero-wavenumber mode is excluded from the summation to avoid division by zero through the $k^2$ denominator. This large-scale forcing, constructed from sinusoidal low-wavenumber modes, has been successfully used in previous FHIT simulations~\cite{abdel2021constant,elghobashi1994predicting,gkoudesnes2020verification,zhiyin2015large}.
This work selects this forcing for two reasons. First, it supplies kinetic energy at large scales, consistent with the conventional energy-input mechanism in FHIT. Second, it is solenoidal: each forced Fourier mode satisfies $\mathbf{k}\cdot\widehat{\mathbf{F}}_{\mathrm{L}}(\mathbf{k})=0$. Therefore, this forcing does not have a compressive component.

\subsection{Real-space composite filtering for extracting middle-scale motions}
\label{sec:Real-space composite filtering}

The construction of intermediate-scale feedback forcing requires a velocity field that selectively represents the target intermediate-scale motions, rather than the instantaneous velocity field including all resolved scales. To this end, we construct a two-stage real-space composite filtering method based on a smoothing operator. In the first stage, the instantaneous velocity field is smoothed repeatedly to obtain a large-scale, low-wavenumber approximation. This approximation is then subtracted from the original velocity field to reduce the low-wavenumber contribution. In the second stage, the residual field is smoothed to suppress high-wavenumber fluctuations, yielding an intermediate-scale velocity field for the subsequent construction of the feedback force. Let \(\mathcal{L}(\cdot)\) denote one local smoothing operation. For an arbitrary scalar field \(q\), the operator is defined using a three-dimensional seven-point stencil as:

\begin{equation}
\mathcal{L}(q_{i,j,k})=\frac{4q_{i,j,k}+q_{i+1,j,k}+q_{i-1,j,k}+q_{i,j+1,k}+q_{i,j-1,k}+q_{i,j,k+1}+q_{i,j,k-1}}{10}.
\label{eq:smoothing_operator}
\end{equation}
Here, \(i\), \(j\) and \(k\) denote the lattice indices in the three spatial directions. And all neighboring nodes are indexed using periodic boundary conditions. The operator assigns a weight of \(4/10\) to the central node and equal weights of \(1/10\) to its six nearest neighbors. Because the weights sum to one, the smoothing operator preserves a spatially uniform field. Repeated application of the operator progressively reduces high-wavenumber fluctuations in the velocity field.

For the velocity vector field \(\mathbf{u}=(u_x,u_y,u_z)^{\mathrm{T}}\), the smoothing operator defined in Eq. \eqref{eq:smoothing_operator} is applied component-wise to the three velocity components, giving the following:
\begin{equation}
\mathcal{L}(\mathbf{u})=\left[\mathcal{L}(u_x),\mathcal{L}(u_y),\mathcal{L}(u_z)\right]^{\mathrm{T}}.
\label{eq:vector_smoothing_operator}
\end{equation}
The local smoothing operator is applied repeatedly for \(r\) iterations to adjust the smoothing strength. This process is denoted by \(\mathcal{L}^{r}\). The iterative process is defined as:
\begin{equation}
\mathcal{L}^{0}(q)=q,\qquad \mathcal{L}^{r}(q)=\mathcal{L}\left[\mathcal{L}^{r-1}(q)\right],\qquad r=1,2,\ldots.
\label{eq:iterative_smoothing_operator}
\end{equation}
As the number of iterations \(r\) increases, the fluctuations of high-wavenumber in the velocity field are smoothed more strongly. Thus, \(\mathcal{L}^{r}\) can be used to construct velocity fields with different degrees of smoothing.
First, the instantaneous velocity field \(\mathbf{u}\) is smoothed for \(n\) iterations to obtain its large-scale approximation:
\begin{equation}
\mathbf{u}_{\mathrm{low}}=\mathcal{L}^{n}(\mathbf{u}).
\label{eq:low_velocity}
\end{equation}
The large-scale approximation is then subtracted from the original velocity field to obtain the residual velocity field:
\begin{equation}
\mathbf{u_\mathrm{res}}=\mathbf{u}-\mathbf{u_\mathrm{low}}=\mathbf{u}-\mathcal{L}^{n}(\mathbf{u}).
\label{eq:residual_velocity}
\end{equation}
This residual velocity field \(\mathbf{u_\mathrm{res}}\) has a reduced contribution to low-wavenumber, but still contains target intermediate-scale motions and higher-wavenumber small-scale fluctuations. It is therefore smoothed for another \(m\) iteration to define the intermediate-scale velocity field used to construct the feedback forcing:
\begin{equation}
\mathbf{u_\mathrm{mid}}=\mathcal{L}^{m}(\mathbf{u_\mathrm{res}})=\mathcal{L}^{m}\left[\mathbf{u}-\mathcal{L}^{n}(\mathbf{u})\right].
\label{eq:mid_velocity}
\end{equation}

Here, \(n\) denotes the number of smoothing iterations in the first stage and is used to construct a large-scale approximation of the instantaneous velocity field. Similarly, \(m\) denotes the number of smoothing iterations in the second stage and is used to attenuate high-wavenumber fluctuations in the residual velocity field. By first removing the large-scale approximation and then smoothing the remaining high-wavenumber components, the procedure yields the intermediate-scale velocity field \(\mathbf{u}_{\mathrm{mid}}\) used to construct the feedback forcing.
Different combinations of \(n\) and \(m\) change the scale composition and amplitude of the extracted velocity field. Several combinations are compared in this work, and the selection criteria and comparison results are presented in Section~\ref{subsec:filter_selection}.

\subsection{Amplitude-limited middle-scale feedback forcing}
\label{sec:middle_scale_feedback}

The intermediate-scale velocity field \(\mathbf{u}_{\mathrm{mid}}\) extracted by the real-space composite filter is used to construct the feedback force. Its local direction follows the filtered velocity, while its magnitude is referenced to the local magnitude of the existing large-scale force. The feedback is intended to strengthen intermediate-scale motions and promote nonlinear energy transfer toward smaller resolved scales.

The resulting intermediate-scale feedback force is defined at each spatial location as
\begin{equation}
\mathbf{F}_{\mathrm{mid}}(\mathbf{x},t)=\alpha |\mathbf{F}_{L}(\mathbf{x},t)|
\frac{\mathbf{u}_{\mathrm{mid}}(\mathbf{x},t)}{|\mathbf{u}_{\mathrm{mid}}(\mathbf{x},t)|+\epsilon_1}.
\end{equation}

Here, \(\alpha\geq0\) is a dimensionless coefficient that controls the feedback strength. The small positive constant \(\epsilon_1\) has the same units as the velocity and prevents division by zero.

Accordingly, the local magnitude of the feedback force satisfies the bound
\begin{equation}
|\mathbf{F}_{\mathrm{mid}}|=\alpha |\mathbf{F}_{L}|\frac{|\mathbf{u}_{\mathrm{mid}}|}
{|\mathbf{u}_{\mathrm{mid}}|+\epsilon_1}\leq\alpha |\mathbf{F}_{L}|.
\end{equation}

For \(|\mathbf{u}_{\mathrm{mid}}|\gg\epsilon_1\), the feedback magnitude is approximately \(\alpha\) times the local large-scale force magnitude. As the filtered velocity approaches zero, the feedback force also vanishes, avoiding finite-amplitude forcing where no clear intermediate-scale direction is available.

By construction, the feedback force has a non-negative coupling with the extracted intermediate-scale velocity:
\begin{equation}\mathbf{u}_{\mathrm{mid}}\cdot\mathbf{F}_{\mathrm{mid}}=
\alpha |\mathbf{F}_{L}|\frac{|\mathbf{u}_{\mathrm{mid}}|^2}{|\mathbf{u}_{\mathrm{mid}}|+\epsilon_1}\geq 0.
\end{equation}

This expression describes the coupling with the filtered velocity used to construct the force. The actual contribution of the feedback to the kinetic energy budget is determined by \(\mathbf{u}\cdot\mathbf{F}_{\mathrm{mid}}\), using the full velocity field.
The above definition specifies the intermediate-scale feedback term \(\mathbf{F}_{\mathrm{mid}}\) introduced in Eq.~(9), which completes the formulation of the composite forcing.
Thus, \(\mathbf{F}_{L}\) retains its role in maintaining turbulence and \(\alpha=0\) recovers the original large-scale forcing case.

\section{Results and Discussion}
\label{sec:results}

\subsection{Selection of the intermediate-scale extraction filter}
\label{subsec:filter_selection}
To construct feedback forcing at the target intermediate scales, we first extract the corresponding velocity component from each instantaneous field. We use the composite physical-space filter defined in Section~\ref{subsec:filter_selection}, based on the iterative smoothing operator $\mathcal{L}$. For a filter configuration denoted by $n+m$, the composite filtering operator is given by Eq.~\eqref{eq:mid_velocity}.
The composite filter is designed to concentrate the spectral energy within the target band while retaining sufficient intermediate-scale response. It also aims to limit low-wavenumber leakage and high-wavenumber contamination. Its design therefore balances spectral concentration and response retention, rather than simply maximizing the total energy of the filtered field.

\begin{figure}[H]
    \centering
    \includegraphics[width=0.6\textwidth]{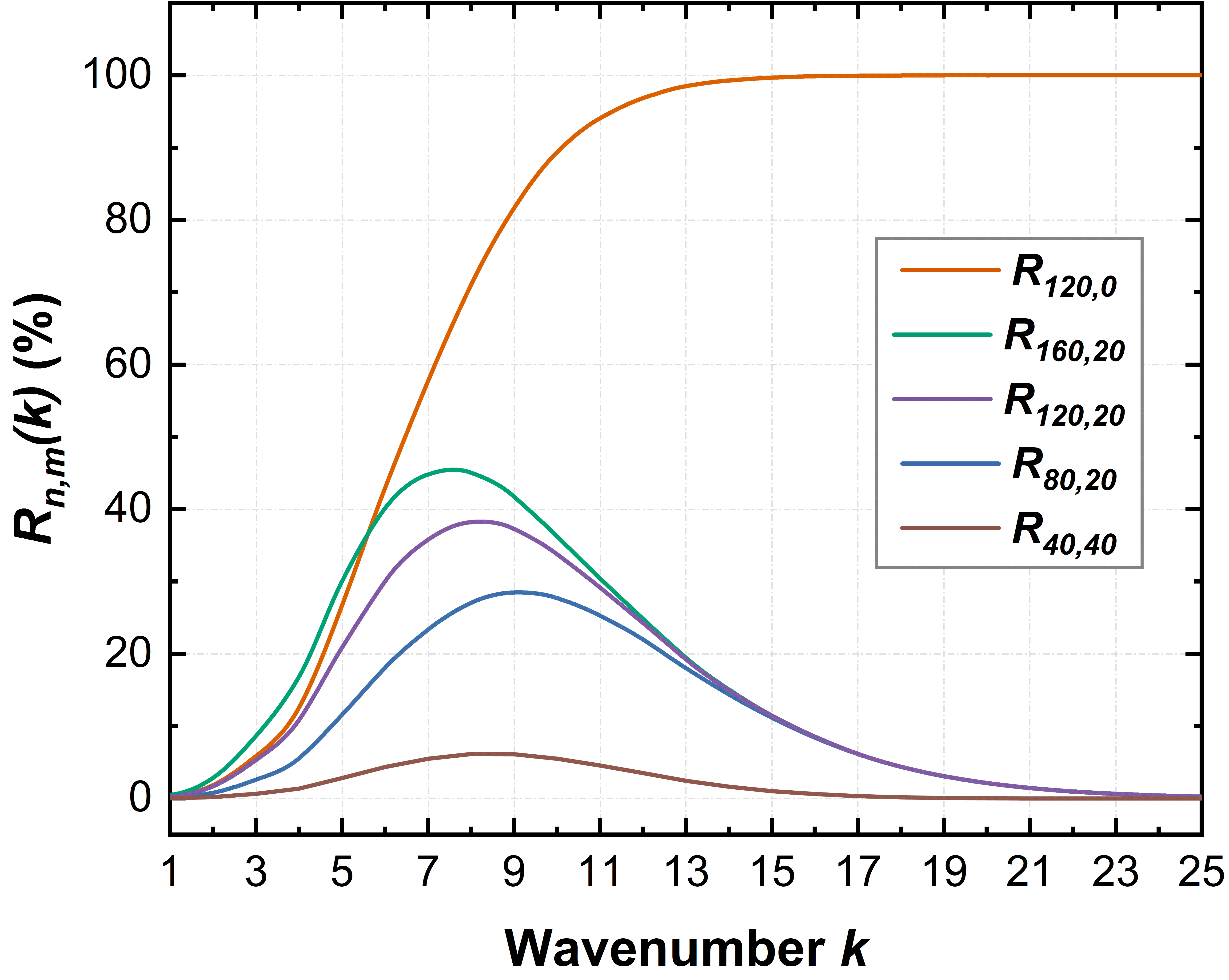}
    \caption{Normalized spectral energy response of the candidate composite filters.}
    \label{fig:filter_response}
\end{figure}

Fig.~\ref{fig:filter_response} shows the normalized spectral response, $R_{n,m}(k)=E_{\mathrm{filtered}}(k)/E_{\mathrm{origin}}(k)$, expressed as a percentage for each candidate filter configuration. For the $120+0$ configuration without outer smoothing, the response increases with wavenumber and approaches $98.50\%$ for $k=13$. Subtracting the smoothed large-scale component therefore reduces low-wavenumber motions while largely preserving high-wavenumber energy in the residual field. This configuration exhibits high-pass characteristics and does not isolate the target intermediate-scale band.

With outer smoothing, the configurations $160+20$, $120+20$, $80+20$, and $40+40$ exhibit band-pass characteristics. Their responses peak at intermediate wavenumbers and decay at higher wavenumbers. The normalized spectral responses of the peak are approximately $45.06\%$, $38.22\%$, $28.49\%$, and $6.15\%$, respectively, indicating substantial differences in response retention among smoothing combinations.
Among candidate configurations, $40+40$ reduces the components of high frequency more highly, but it also produces a substantially lower response across the target intermediate-scale band. Assessing filter performance requires consideration of both response strength and scale selectivity. We therefore also examine the relative distribution of filtered energy among the low-wavenumber, target intermediate-scale, and high-wavenumber bands.

To further quantify the scale selectivity of the filtering configurations, we divide the spectral energy of the filtered velocity field into three wavenumber bands: the low-wavenumber band $k<5$, the target intermediate-scale band $5\leq k\leq12$, and the high-wavenumber band $k>12$. The energy in each band, expressed as a fraction of the total filtered energy, is defined as:
\begin{equation}
\begin{aligned}
C_{\mathrm{low}}^{(n,m)}&=\frac{\displaystyle\sum_{1\leq k<5}
E_{\mathrm{filtered}}^{(n,m)}(k)}{\displaystyle\sum_{k\geq1}
E_{\mathrm{filtered}}^{(n,m)}(k)}\times 100\%,
C_{\mathrm{target}}^{(n,m)}&=\frac{\displaystyle\sum_{5\leq k\leq12}
E_{\mathrm{filtered}}^{(n,m)}(k)}{\displaystyle\sum_{k\geq1}
E_{\mathrm{filtered}}^{(n,m)}(k)}\times 100\%,
C_{\mathrm{high}}^{(n,m)}&=\frac{\displaystyle\sum_{k>12}
E_{\mathrm{filtered}}^{(n,m)}(k)}{\displaystyle\sum_{k\geq1}
E_{\mathrm{filtered}}^{(n,m)}(k)}\times 100\%.
\end{aligned}
\label{eq:filter_energy_fractions}
\end{equation}
Here, $C_{\mathrm{target}}$ characterizes the concentration of filtered energy within the target intermediate-scale band, while $C_{\mathrm{low}}$ and $C_{\mathrm{high}}$ characterize the relative levels of low-wavenumber leakage and high-wavenumber contamination, respectively.

\begin{figure}[H]
    \centering
\includegraphics[width=0.6\textwidth]{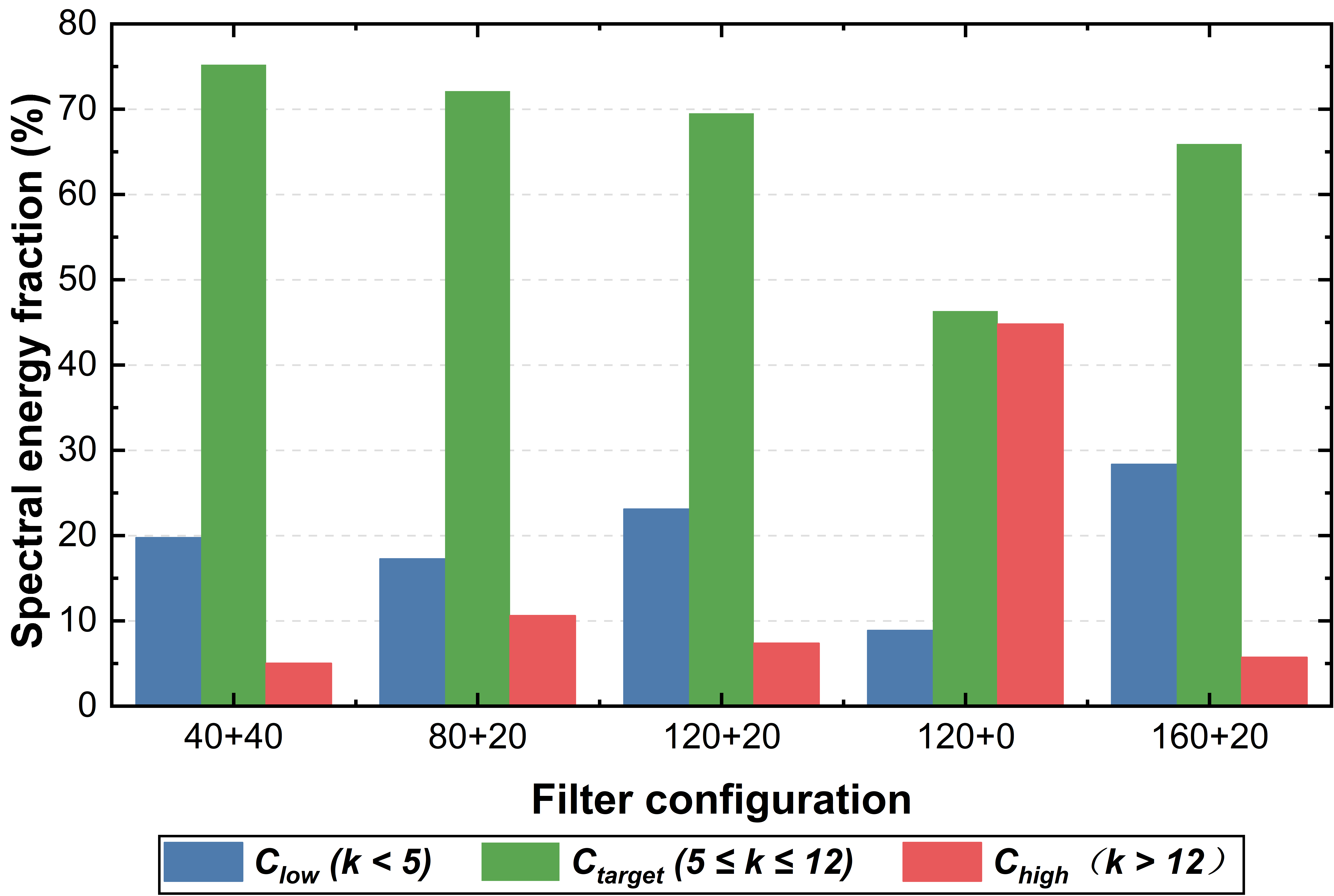}
    \caption{Band-wise spectral energy fractions of the filtered velocity fields for the candidate composite filters.}
    \label{fig:filter_energy_partition}
\end{figure}

Fig.~\ref{fig:filter_energy_partition} shows that $40+40$ has the highest target-band energy fraction ($75.23\%$) and the lowest high-wavenumber fraction ($5.03\%$) among the candidate configurations. Therefore, it provides the greatest relative concentration of filtered energy within the target band. By contrast, the $120+0$ configuration retains $44.79\%$ of its filtered energy in the high-wavenumber band. Removing the large-scale component without outer smoothing thus leaves a substantial high-wavenumber contribution in the residual field, limiting its selectivity for the target intermediate scales.
However, the high target-band fraction of $40+40$ does not imply greater energy retention at the target scales. As shown in Fig.~\ref{fig:filter_response}, its normalized spectral response within the target band is substantially lower than those of $80+20$, $120+20$, and $160+20$. Its greater relative energy concentration therefore comes with stronger attenuation of the target-scale response, highlighting the need to assess both concentration and response retention.

For configurations with $20$ outer smoothing iterations, increasing the number of inner iterations introduces a different trade-off. The high-wavenumber fractions of $120+20$ and $160+20$ are $7.35\%$ and $5.72\%$, respectively, compared to $10.62\%$ for $80+20$. Their low-wavenumber fractions increase to $23.11\%$ and $28.35\%$, exceeding $17.25\%$ for $80+20$. Since large-scale forcing that maintains turbulence primarily acts at low wavenumbers, limiting low wavenumber leakage helps reduce potential interference. Therefore, filter selection should focus more on limiting low-wavenumber leakage. Compared with the two configurations above, $80+20$ maintains a target-band energy fraction of $72.13\%$ while retaining a lower low-wavenumber energy fraction. It better achieves the design objective of concentrating the feedback at intermediate scales.

Taking into account the normalized spectral response, band-wise energy distribution, and filtering cost, $80+20$ offers a practical compromise among candidate configurations. It has neither the highest target-band energy fraction nor the lowest high-wavenumber fraction. However, it retains a stronger response at the target-scale than $40+40$, with $72.13\%$ of the filtered energy concentrated within $5\leq k\leq12$. It also has the lowest low-wavenumber energy fraction among these three configurations and requires fewer smoothing iterations than $120+20$ and $160+20$. Therefore, this work selects $80+20$ for intermediate-scale velocity extraction, balancing response retention, scale selectivity, and computational cost.

\subsection{Intermediate-scale feedback power and spectral distribution}
\label{subsec:midforce_verification}

Having selected the intermediate-scale velocity extraction method, this section examines the power response and scale selectivity of the resulting feedback. The feedback force is constructed from the filtered velocity $\mathbf{u}_{\mathrm{mid}}$, rather than the full velocity field. It is aligned with the filtered intermediate-scale velocity. The magnitude of their cross-power is expected to be concentrated primarily within the prescribed target band.
Firstly, examine the spatially averaged construction power $P_{\mathrm{mid}}$ and its response to the feedback coefficient $\alpha$:
\begin{equation}
P_{\mathrm{mid}}=\left\langle\mathbf{u}_{\mathrm{mid}}\cdot\mathbf{F}_{\mathrm{mid}}\right\rangle_{\Omega},
\label{eq}
\end{equation}
where $\langle\cdot\rangle_{\Omega}$ denotes spatial averaging in the computational domain $\Omega$. Then we use the cross-power spectrum $P_c(k)$ and its band-wise fractions to characterize the spectral distribution of the feedback--velocity coupling and the temporal persistence of its target-band concentration.

\begin{figure}[H]
    \centering
\includegraphics[width=0.6\textwidth]{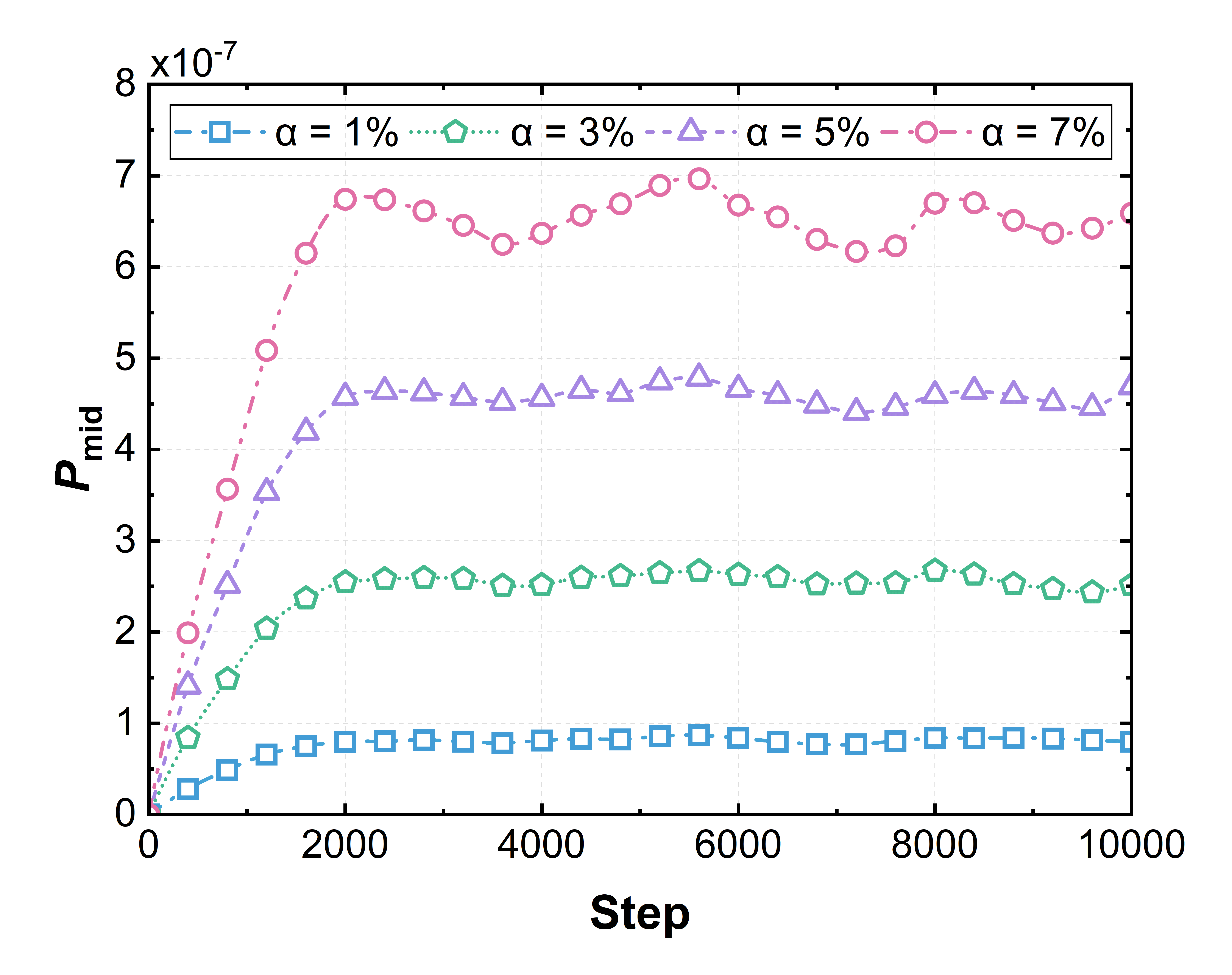}
\caption{Temporal evolution of the spatially averaged feedback–filtered-velocity coupling power \(P_{\mathrm{mid}}\) for different feedback coefficients \(\alpha\).}
    \label{fig:UF}
\end{figure}

As shown in Fig.~\ref{fig:UF}, \(P_{\mathrm{mid}}\) remains positive after the initial transient for all nonzero feedback strengths. For \(\alpha=1\%\), \(3\%\), \(5\%\), and \(7\%\), its time-averaged values over the stationary sampling interval are \(0.75\times10^{-7}\), \(2.53\times10^{-7}\), \(4.64\times10^{-7}\), and \(6.41\times10^{-7}\), respectively, with each curve fluctuating around its corresponding mean. The quasi-steady level of \(P_{\mathrm{mid}}\) increases with \(\alpha\), and the order of the curves remains unchanged, indicating that the feedback coefficient $\alpha$ can control the construction power.

To further examine whether the feedback-velocity coupling is concentrated primarily within the target band, we compute the shell-integrated cross-power spectrum between the target intermediate-scale velocity field and the feedback force to quantify the contribution of each wavenumber shell to the coupling. The cross-power spectrum is defined as:
\begin{equation}
P_c(k,t)=\sum_{\boldsymbol{\kappa}\in\mathcal{S}_k}\operatorname{Re}\left[\widehat{\mathbf{u}}_{\mathbf{mid}}^{*}(\boldsymbol{\kappa},t)\cdot\widehat{\mathbf{F}}_{\mathrm{mid}}(\boldsymbol{\kappa},t)\right].
\label{eq}
\end{equation}
Here, \(\mathcal{S}_k\) denotes the set of all Fourier modes within the spherical shell indexed by \(k\). The quantities \(\widehat{\mathbf{u}}_{\mathrm{mid}}\) and \(\widehat{\mathbf{F}}_{\mathrm{mid}}\) are the Fourier coefficients of the target intermediate-scale velocity field and the intermediate-scale feedback force, respectively. The superscript \(*\) denotes complex conjugation, and \(\operatorname{Re}(\cdot)\) denotes the real part.
To quantify the spectral concentration of the feedback-velocity coupling, we define the target-band fraction as:
\begin{equation}
C_{\mathrm{target}}^{P}(t)=\frac{\sum_{5\leq k\leq12}|P_c(k,t)|}{\sum_k|P_c(k,t)|}\times100\%
\label{eq}
\end{equation}
The absolute value \(|P_c(k,t)|\) is used to quantify the magnitude of the coupling contributions across different wavenumber ranges while preventing positive and negative contributions from canceling during summation. Thus, \(C_{\mathrm{target}}^{P}\) characterizes the spectral concentration of the coupling within the target band, rather than the algebraic distribution of the net power across scales.

\begin{figure}[h]
    \centering
    \includegraphics[width=0.6\textwidth]{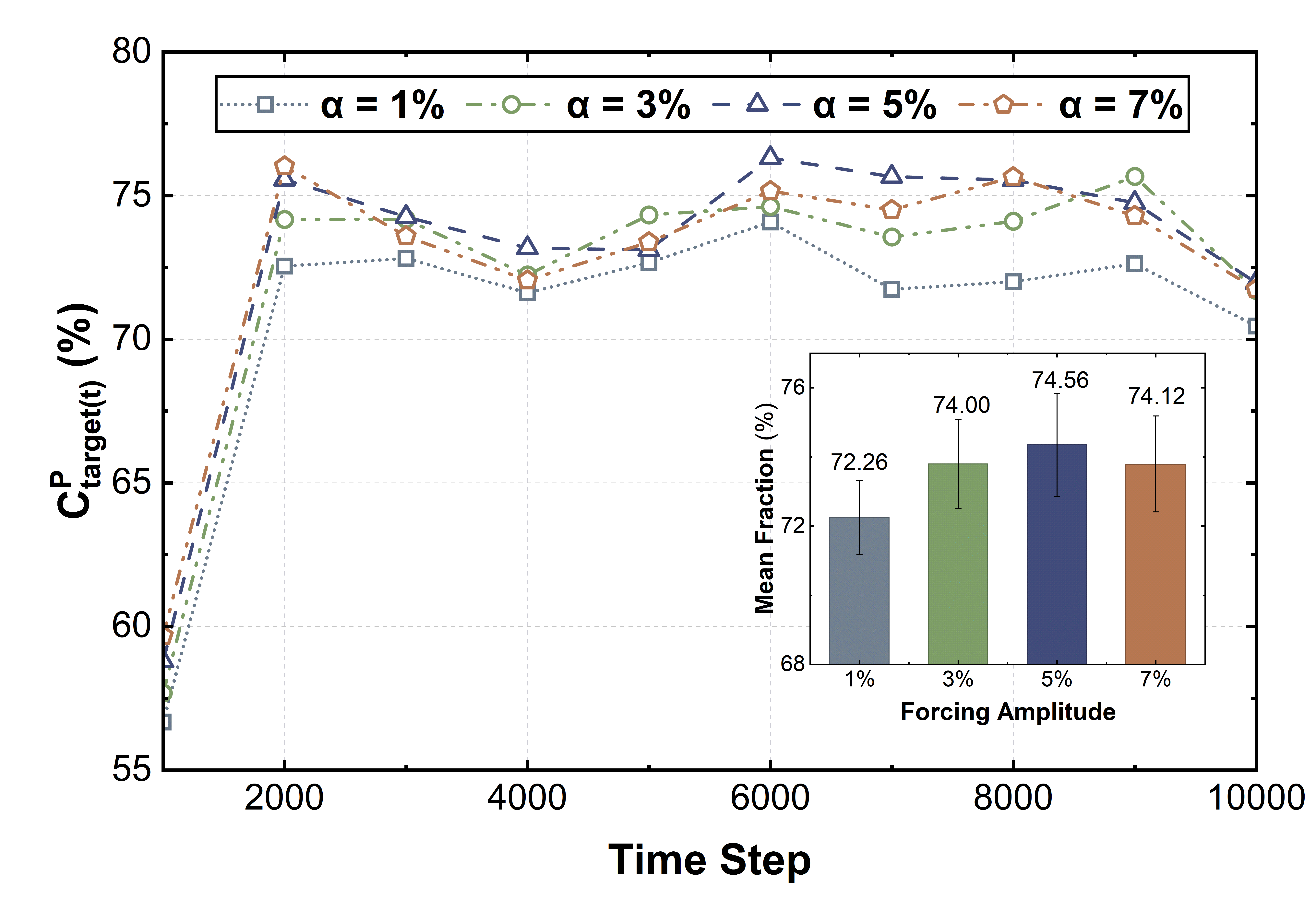}
    \caption{Temporal evolution of the target-band fraction \(C_{\mathrm{target}}^{P}(t)\) of the cross-power magnitude \(|P_c(k,t)|\) for different feedback coefficients \(\alpha\). Inset: steady-state temporal means with standard-error bars.}
    \label{fig:SP}
\end{figure}

As shown in Fig.~\ref{fig:SP}, the target-band fractions for the different feedback strengths $\alpha$ range from approximately \(56.67\%\) to \(59.68\%\) during the initial development stage. As the flow develops, these fractions increase rapidly and remain within the range of \(70.46\%\text{--}76.31\%\). This indicates that the dominant spectral contribution to the coupling remains concentrated within the prescribed target band, with no dominance of contributions from outside this band. The feedback constructed from the composite-filtered velocity therefore maintains a consistent scale selectivity.

\begin{figure}[h]
    \centering
    \includegraphics[width=0.6\textwidth]{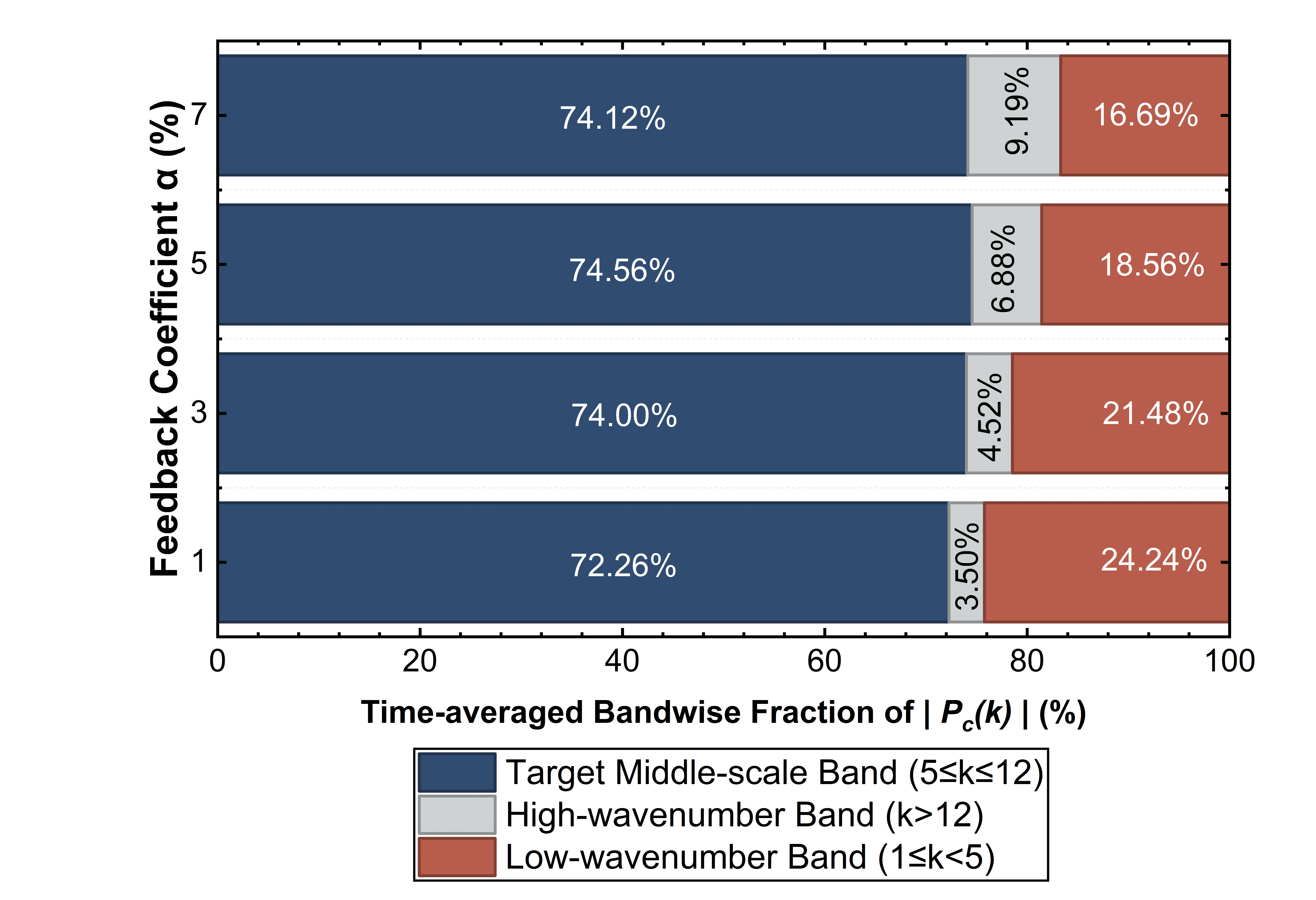}
    \caption{Time-averaged distribution of the cross-power magnitude \(\left|P_c(k,t)\right|\) among the low-wavenumber, target intermediate-scale, and high-wavenumber bands for different nonzero feedback coefficients \(\alpha\).}
    \label{fig:FC}
\end{figure}

To further examine the distribution of coupling across wavenumber ranges, we use the cross-power spectral magnitude \(\left|P_c(k)\right|\) to calculate the spectral magnitude fractions in the low-wavenumber, target intermediate-scale and high-wavenumber bands and then average these fractions over the stationary sampling interval.
As shown in Fig.~\ref{fig:FC}, the target intermediate-scale band accounts for the dominant fraction of the cross-power spectral magnitude \(\left|P_c(k)\right|\) in all nonzero feedback strengths. Its time-averaged fraction ranges from \(72.26\%\) to \(74.56\%\), substantially exceeding those of the low- and high-wavenumber bands. These results further demonstrate the scale selectivity of the constructed intermediate-scale feedback--velocity coupling, with its dominant spectral magnitude contribution concentrated within the prescribed target band.

As the feedback coefficient $\alpha$ increases from 1\% to 5\%, the target-band fraction increases slightly from \(72.26\%\) to \(74.56\%\), while the low-wavenumber fraction decreases from \(24.24\%\) to \(18.56\%\) and the high-wavenumber fraction increases from \(3.50\%\) to \(6.88\%\). Over this range of feedback strengths, the low-wavenumber contribution to the spectral magnitude decreases in relative terms, accompanied by a slight increase in target-band concentration and a larger relative contribution at high wavenumbers. 
When the feedback coefficient $\alpha$ is further increased to 7\%, the target-band fraction decreases slightly to $74.12\%$, while the high-wavenumber fraction increases to $9.19\%$. Thus, the case $\alpha=5\%$ exhibits the best scale selectivity among the cases examined. Further increasing the feedback does not improve target-band concentration but increases the relative contribution of non-target high-wavenumber components to the cross-power spectral magnitude. This indicates that stronger feedback does not necessarily yield better scale selectivity.

In summary, in all nonzero feedback strengths examined, \(P_{\mathrm{mid}}\) remained positive after the initial transient, and its quasi-steady mean increased with \(\alpha\). The magnitude of the cross-power spectrum between the feedback force and the filtered velocity remained mainly concentrated within the target band. We next examine the resulting changes in energy spectra and band-integrated energies.

\subsection{Spectral energy response to intermediate-scale feedback}
Although the constructed intermediate-scale feedback--velocity coupling exhibits the intended power response and clear scale selectivity. However, nonlinear energy transfer between scales may extend the energy response to the feedback beyond the target band. This section examines kinetic energy spectra, band-integrated energies, and the time evolution of energy fractions at different feedback strengths to assess how intermediate-scale feedback changes the flow's energy distribution.

\begin{figure}[H]
    \centering
    \includegraphics[width=0.6\textwidth]{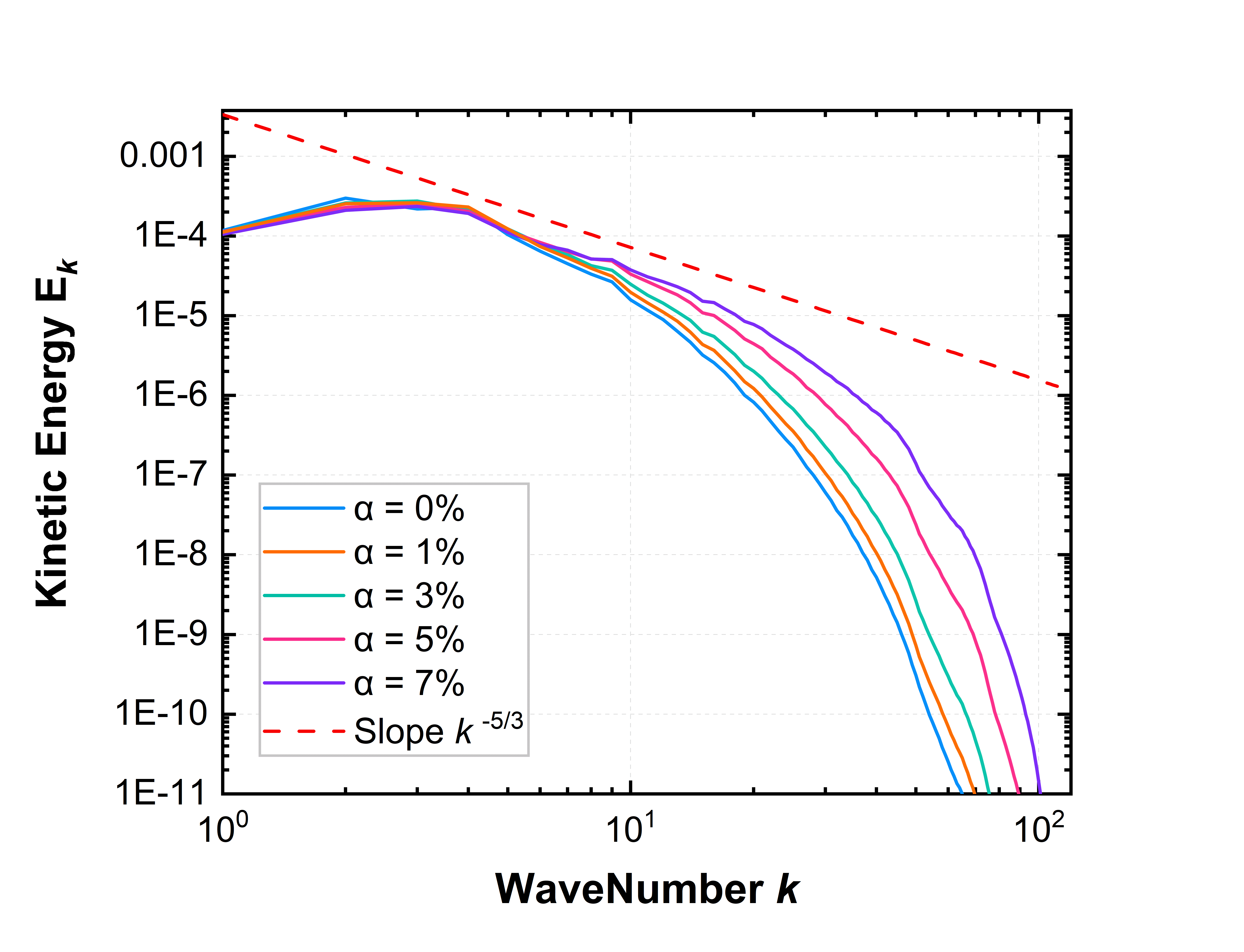}
\caption{Kinetic energy spectra \(E(k)\) under different middle-scale feedback coefficient \(\alpha\). The dashed line indicates the \(k^{-5/3}\) scaling for reference.}
    \label{fig:Sprctura}
\end{figure}

Fig.~\ref{fig:Sprctura} shows the time-averaged kinetic energy spectra \(E(k)\) at different feedback coefficients \(\alpha\). At low wavenumbers, the curves remain relatively close, with comparatively small differences in the energy distribution. Within the target band \(5\leq k\leq12\), the curves separate and the spectral energy generally increases with the feedback coefficient \(\alpha\). The differences are more evident in the high wavenumber range, where a stronger feedback coefficient \(\alpha\) produces higher spectral energy levels and more extended spectral tails.

\begin{figure}[H]
    \centering
    \includegraphics[width=0.6\textwidth]{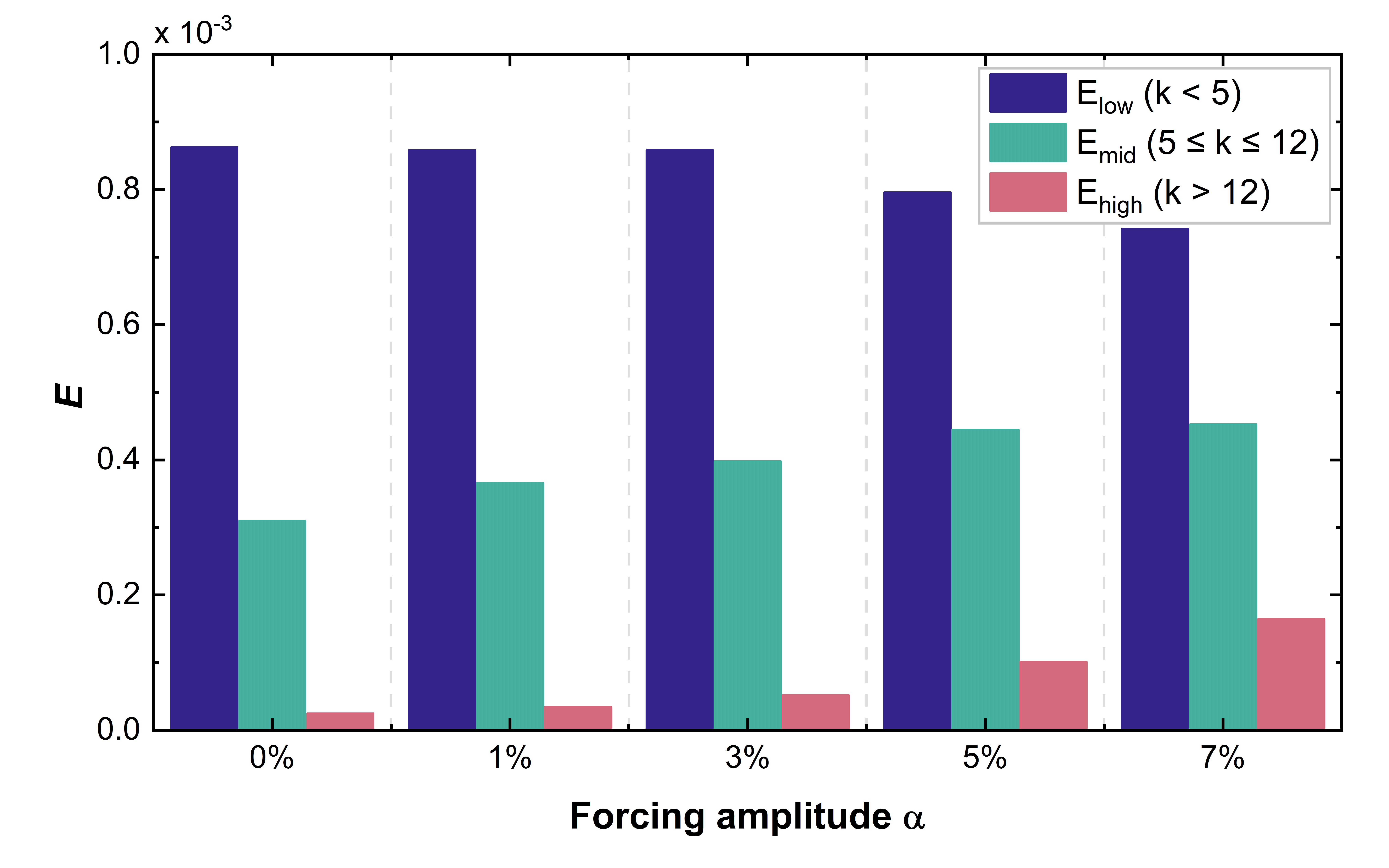}
\caption{Band-integrated kinetic energies in the low-wavenumber (\(k<5\)), target intermediate-scale (\(5\leq k\leq12\)), and high-wavenumber (\(k>12\)) ranges for different intermediate-scale feedback coefficients \(\alpha\).}
    \label{fig:Total_E}
\end{figure}

To quantify the energy response across scales, we divide the kinetic energy spectrum into three bands: low wavenumbers (\(k<5\)), target intermediate scales (\(5\leq k\leq12\)), and high wavenumbers (\(k>12\)). We calculate the corresponding band-integrated energies \(E_{\mathrm{low}}\), \(E_{\mathrm{mid}}\), and \(E_{\mathrm{high}}\). As shown in Fig.~\ref{fig:Total_E}, the low-wavenumber energy remains close to the baseline level under weaker feedback but obviously decreases at \(\alpha=5\%\) and \(7\%\). In contrast, the intermediate-scale energy increases with the feedback coefficient $\alpha$, from approximately \(3.10\times10^{-4}\) in the baseline case to \(4.53\times10^{-4}\) in \(\alpha=7\%\). The energy of the high wavenumber range increases from approximately \(2.48\times10^{-5}\) to \(1.64\times10^{-4}\), representing a more pronounced relative increase. In particular, increasing \(\alpha\) from \(5\%\) to \(7\%\) yields only a limited additional gain in intermediate-scale energy, while the high-wavenumber energy continues to increase substantially. Stronger feedback produces an increasingly pronounced energy response at higher wavenumbers, beyond the target intermediate-scale band.

\begin{figure}[H]
    \centering
    \includegraphics[width=0.6\textwidth]{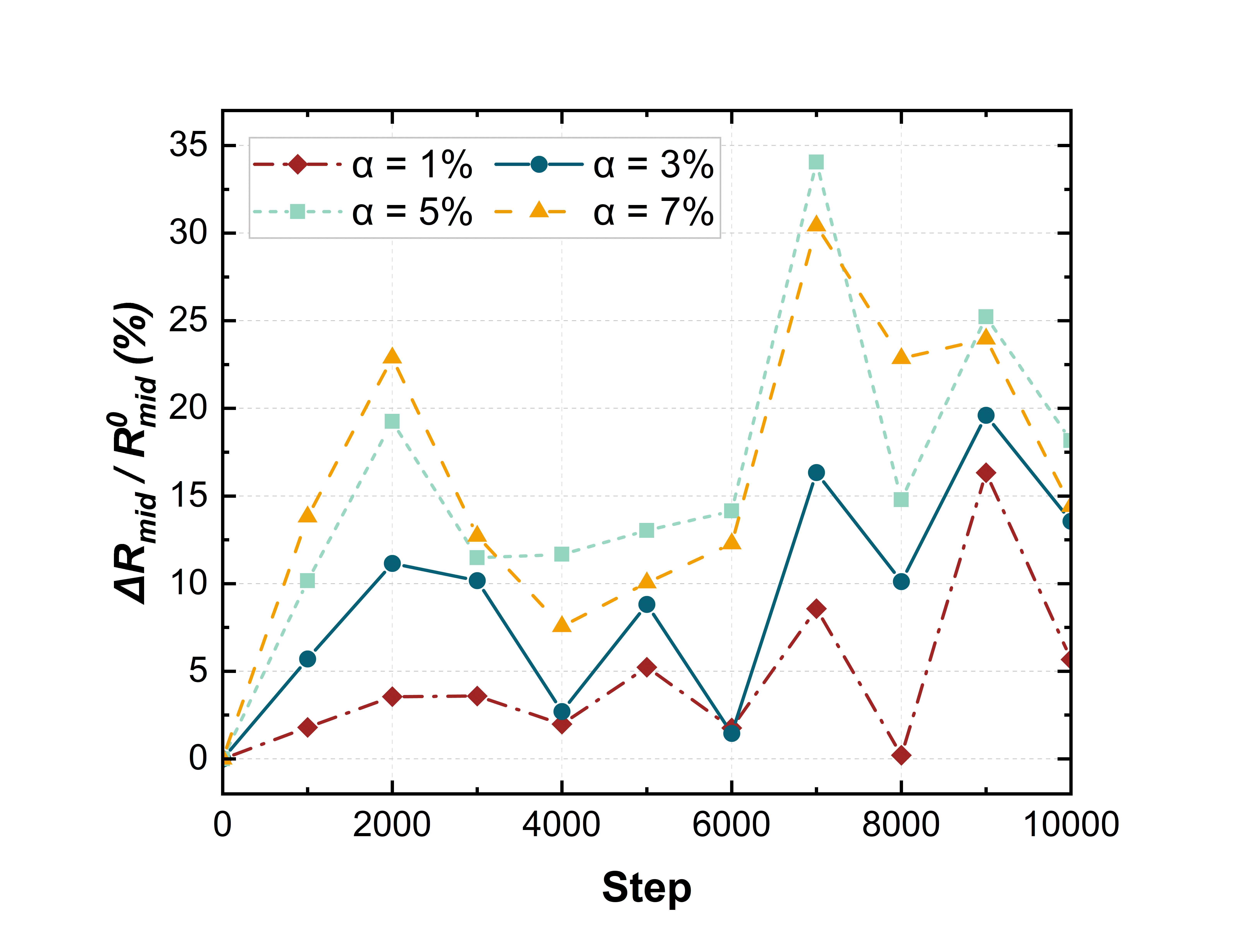}
\caption{Temporal evolution of the relative change in the target intermediate-scale energy fraction, \(\Delta R_{\mathrm{mid}}/R_{\mathrm{mid}}^{0}\), for different intermediate-scale feedback coefficients \(\alpha\), relative to the baseline case.}
    \label{fig:Rmid}
\end{figure}

To characterize the relative distribution of kinetic energy across wavenumber ranges, we define the target intermediate-scale and high-wavenumber energy fractions as:
\begin{equation}
R_{\mathrm{mid}}^{\alpha}(t)=\frac{E_{\mathrm{mid}}^{\alpha}(t)}
{E_{\mathrm{total}}^{\alpha}(t)},
\qquad
R_{\mathrm{high}}^{\alpha}(t)=\frac{E_{\mathrm{high}}^{\alpha}(t)}
{E_{\mathrm{total}}^{\alpha}(t)}.
\end{equation}
Using the case of no feedback \(\alpha=0\%\), as a baseline, we define the percentage change in each band-energy fraction as:
\begin{equation}
\frac{\Delta R_b^{\alpha}(t)}{R_b^{0}(t)}=\frac{R_b^{\alpha}(t)-R_b^{0}(t)}{R_b^{0}(t)}\times100\%,
\qquad
b=\mathrm{mid},\mathrm{high}.
\end{equation}
Fig.~\ref{fig:Rmid} shows \(\Delta R_{\mathrm{mid}}/R_{\mathrm{mid}}^{0}\) in different feedback coefficients. The relative changes are positive at most sampled times after the initial transient, indicating that the increase in the target-band fraction is not confined to a single sampled instant. During the stationary sampling interval, the mean relative increases are approximately \(6.29\%\), \(11.65\%\), \(18.99\%\), and \(19.89\%\) for \(\alpha=1\%\), \(3\%\), \(5\%\), and \(7\%\), respectively. Therefore, the mean relative increases with the feedback coefficient from \(\alpha=1\%\) to \(5\%\). However, the mean responses at \(5\%\) and \(7\%\) are comparable, and further increasing the feedback strength does not give a higher mean energy fraction response. All cases exhibit temporal fluctuations in the response of the relative energy fraction, indicating that the instantaneous response varies over time.

\begin{figure}[H]
    \centering
    \includegraphics[width=0.6\textwidth]{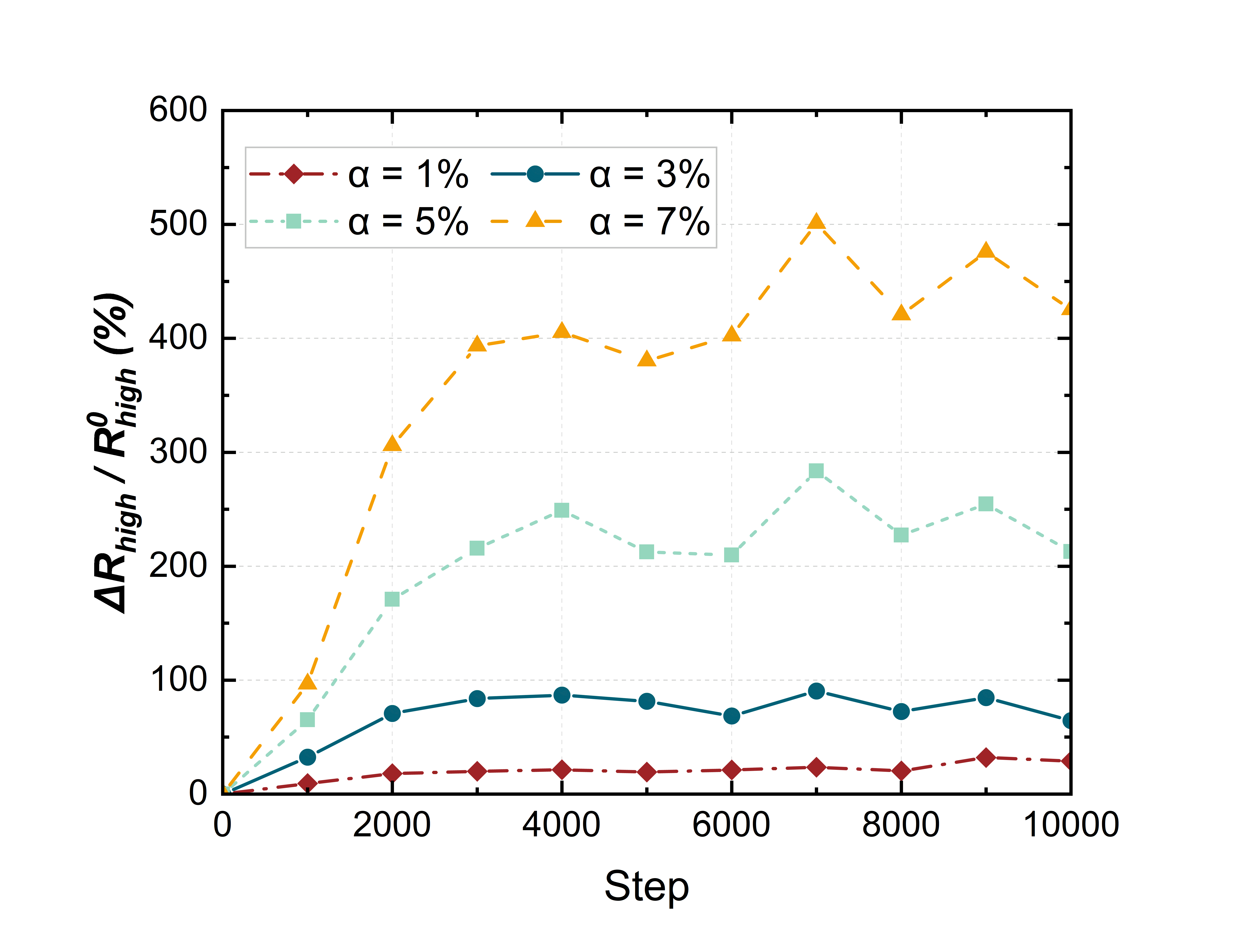}
\caption{Temporal evolution of the relative change in the high-wavenumber energy fraction, \(\Delta R_{\rm high}/R_{\rm high}^{0}\), for different intermediate-scale feedback coefficients \(\alpha\), relative to the baseline case.}
    \label{fig:Rhigh}
\end{figure}

Fig.~\ref{fig:Rhigh} shows the time evolution of \(\Delta R_{\mathrm{high}}/R_{\mathrm{high}}^{0}\) at different feedback coefficients to characterize the temporal response of the high-wavenumber energy fraction. After the initial transient, all feedback cases exhibit positive relative changes, with magnitudes that increase markedly with the feedback coefficient \(\alpha\). After the stationary sampling interval, the mean relative increases are approximately \(24.27\%\), \(77.01\%\), \(233.46\%\), and \(434.29\%\) for \(\alpha=1\%\), \(3\%\), \(5\%\), and \(7\%\), respectively. Whereas the mean relative responses of the intermediate-scale energy fraction are comparable at \(5\%\) and \(7\%\), the high-wavenumber response continues to increase markedly.  These results indicate that changes in the relative energy distribution become increasingly pronounced in the high-wavenumber range as feedback strengthens.

In summary, intermediate-scale feedback increased energy in both the intermediate-scale and high-wavenumber bands, with a larger relative increase in the latter. However, a higher high-wavenumber energy alone does not demonstrate a stronger forward nonlinear energy transfer. We therefore examine the nonlinear energy flux \(\Pi(K)\) to assess net forward energy transfer across wavenumber boundaries.

\subsection{Nonlinear cross-scale energy transfer under intermediate-scale feedback}

The intermediate-scale feedback clearly modifies the energy distribution of the flow: energy within the intermediate-scale band increases, while the high-wavenumber energy response becomes more evident under stronger feedback. However, energy spectra and band-integrated energies cannot verify whether these changes are along with stronger net forward nonlinear energy transfer. To examine this, we analyze the nonlinear energy transfer spectrum \(T(k,t)\) and the corresponding nonlinear energy flux \(\Pi(K,t)\), defined as:
\begin{equation}
\Pi(K,t)=-\sum_{k=0}^{K}T(k,t).
\end{equation}
\(\Pi(K,t)>0\) shows nonlinear energy transfer from larger scales (\(k\leq K\)) to smaller scales (\(k>K\)). To characterize the mean transfer under statistically stationary conditions, we define the time-averaged cumulative nonlinear energy flux as:
\begin{equation}
\overline{\Pi}(K)=\frac{1}{N_t}\sum_{n=1}^{N_t}\Pi(K,t_n),
\end{equation}
where \(t_n\) denotes the \(n\) sampling time within the statistical interval, and \(N_t\) is the total number of sampling times.

\begin{figure}[H]
    \centering
    \includegraphics[width=0.6\textwidth]{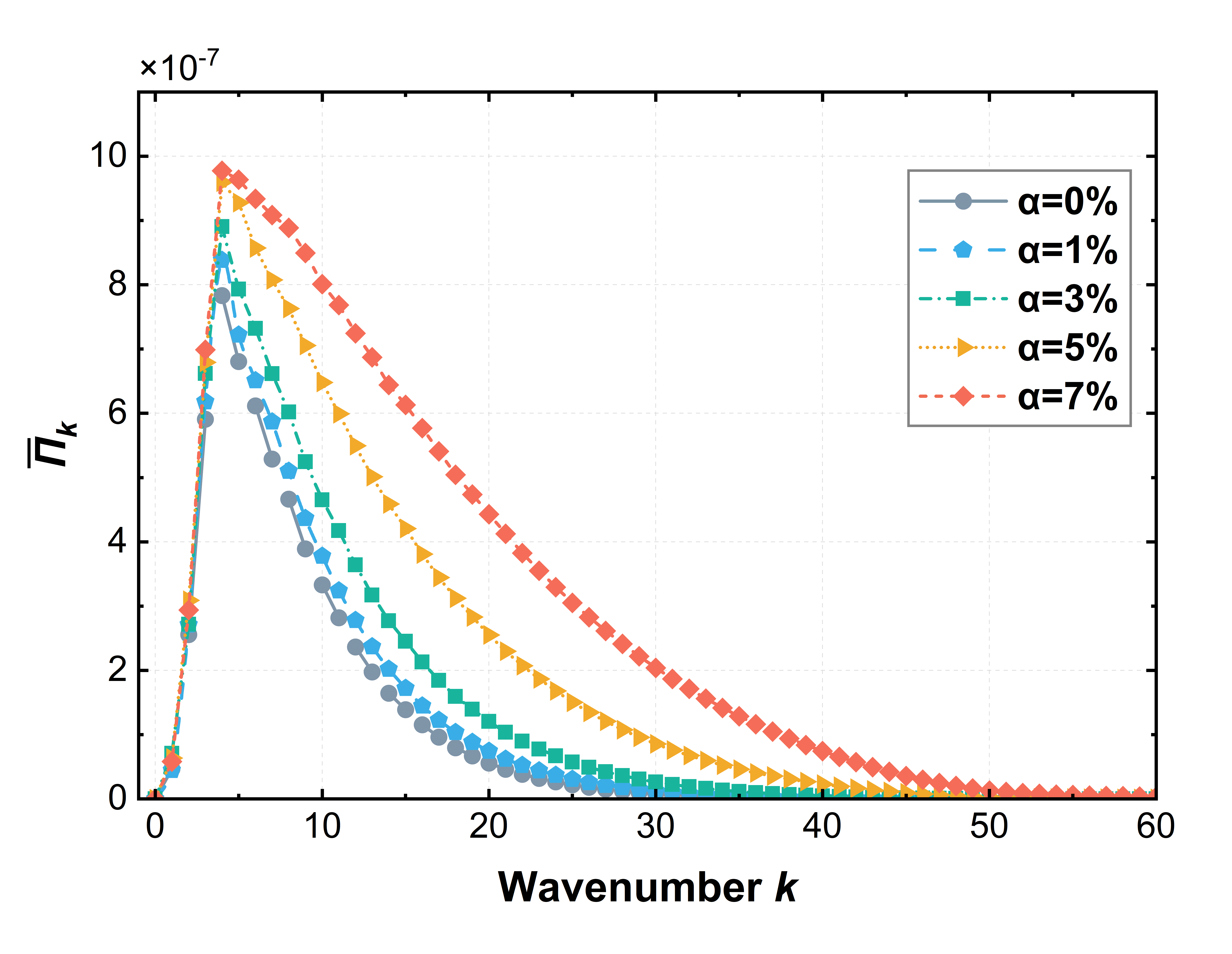}
\caption{Time-averaged cumulative nonlinear energy flux \(\overline{\Pi}(k)\) for different feedback coefficients \(\alpha\). Positive values indicate net forward nonlinear energy transfer toward higher wavenumbers.}
    \label{fig:k}
\end{figure}

As shown in Fig.~\ref{fig:k}, the time-averaged nonlinear energy flux \(\overline{\Pi}(K)\) remains positive over the main wavenumber range at all feedback coefficients, indicating net forward nonlinear energy transfer from larger to smaller scales. In all cases, the peaks near \(K\approx4\) and then gradually decreases with increasing wavenumber. As the feedback coefficient \(\alpha\) increases, the nonlinear energy flux generally increases, while the peak location remains approximately unchanged. Compared with the no-feedback baseline, stronger feedback sustains higher flux levels at intermediate and high wavenumbers beyond the peak. The nonlinear energy flux then gradually approaches zero. The degree of enhancement varies with wavenumber, indicating that the response to feedback from transfer is scale dependent.

\begin{figure}[H]
    \centering
    \includegraphics[width=0.6\textwidth]{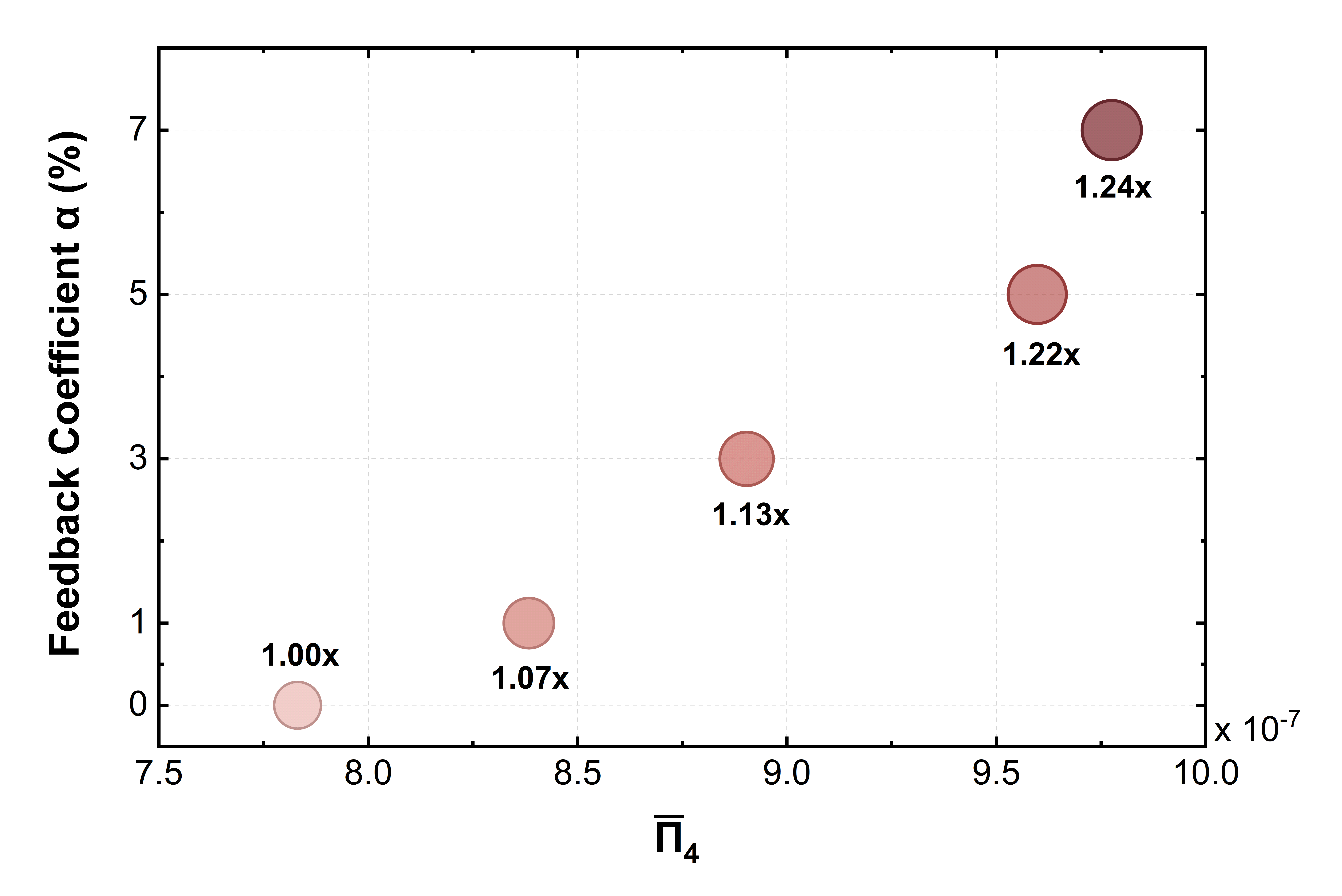}
\caption{Time-averaged cumulative nonlinear energy flux \(\overline{\Pi}(4)\) across the low-wavenumber boundary of the target intermediate-scale band for different feedback coefficients \(\alpha\). The labels indicate the flux magnitude relative to the no-feedback case.}
    \label{fig:4}
\end{figure}

\begin{figure}[H]
    \centering
    \includegraphics[width=0.6\textwidth]{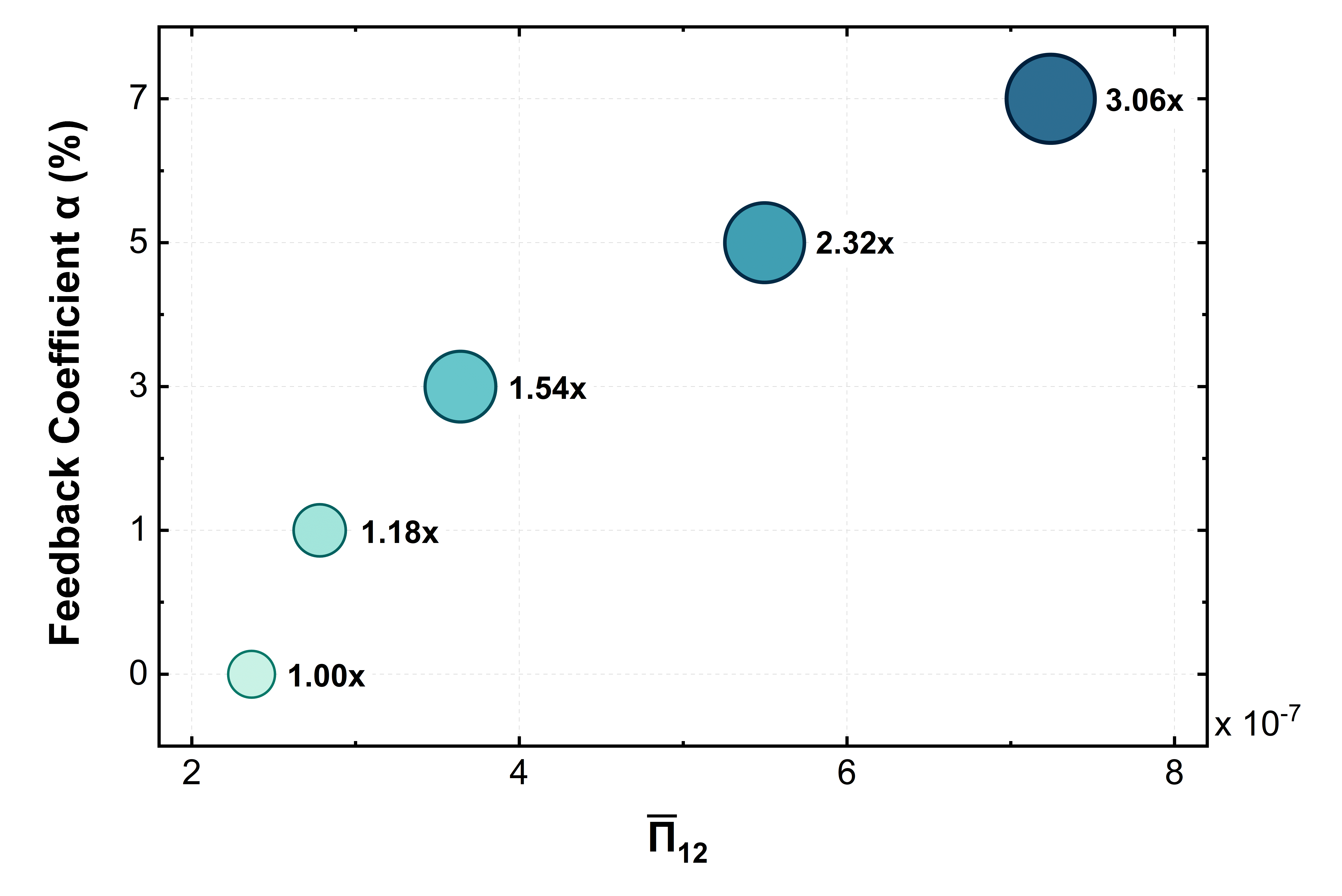}
\caption{Time-averaged cumulative nonlinear energy flux \(\overline{\Pi}(12)\) across the high-wavenumber boundary of the target intermediate-scale band for different feedback coefficients \(\alpha\). The labels indicate the flux magnitude relative to the no-feedback case.}
    \label{fig:12}
\end{figure}

To quantify nonlinear energy transfer across the low-wavenumber boundary of the target intermediate-scale band, we examine \(\overline{\Pi}(4)\) at different feedback coefficients. As shown in Fig.~\ref{fig:4}, with \(\alpha\) increasing from \(0\%\) to \(7\%\), \(\overline{\Pi}(4)\) increasing from \(7.83\times10^{-7}\) to \(9.78\times10^{-7}\), reaching approximately \(1.24\) times the baseline without feedback. This indicates that intermediate-scale feedback enhances the net forward nonlinear energy flux across \(K=4\). However, increasing \(\alpha\) from \(5\%\) to \(7\%\) shows only a further increase of approximately \(1.86\%\) in \(\overline{\Pi}(4)\), which is compared to the baseline without feedback. Therefore, the additional nonlinear energy flux gain at the low-wavenumber boundary is limited over this range.
To quantify nonlinear energy transfer across the boundary of a high-wavenumber range in the intermediate-scale band, we examine \(\overline{\Pi}(12)\) at different feedback coefficients. As shown in Fig.~\ref{fig:12}, $\overline{\Pi}(12)$ increases from $2.36\times10^{-7}$ to $7.24\times10^{-7}$ as $\alpha$ increases from $0\%$ to $7\%$. At $\alpha=7\%$, the mean nonlinear energy flux is approximately $3.06$ times that of the baseline. This indicates that intermediate-scale feedback markedly enhances the net forward nonlinear energy flux across \(K=12\).

The different nonlinear energy flux responses at the two boundaries indicate that the enhancement of cross-scale energy transfer by intermediate-scale feedback is scale dependent. The nonlinear energy flux across $K=4$ shows a smaller relative increase than that across $K=12$. Transfer from large scales to higher-wavenumber regions therefore shows a limited relative response, whereas net forward nonlinear transfer into the small-scale region $k>12$ is enhanced more strongly. Together with the previously observed increase in target-band energy, these findings support the strategy's ability to selectively regulate intermediate-scale energy distribution and cross-scale transfer. This behavior is also consistent with the retention of large-scale forcing, while focusing the feedback on intermediate scales. These nonlinear energy flux results provide further support for the enhancement of the forward energy cascade.

\subsection{Feedback coefficient sensitivity of flow response and numerical applicability}

Increasing the feedback coefficient \(\alpha\) may also change the state of turbulence and local flow statistics. Stronger small-scale activity requires higher numerical resolution. Therefore, we examine how the feedback coefficient \(\alpha\) affects both the physical flow response and the numerical applicability. First, we use the Taylor microscale Reynolds number \(Re_{\lambda}\) and enstrophy to characterize changes in global flow characteristics and vorticity activity. We then examine skewness and the joint \(Q\)--\(R\) distribution to assess changes in velocity-gradient structure. Finally, we use \(k_{\max}\eta\) to assess small-scale spatial resolution.

To characterize the state of turbulence at different feedback coefficients, we examine the Taylor microscale Reynolds number \(Re_\lambda\), defined as:
\begin{equation}
Re_\lambda=\frac{\textbf{u}_{\mathrm{rms}}\lambda}{\nu}, \qquad \lambda=\sqrt{\frac{15\nu \textbf{u}_{\mathrm{rms}}^{2}}{\varepsilon}},
\end{equation}
where \(\textbf{u}_{\mathrm{rms}}\) is the fluctuation in the root-mean-square velocity, calculated as:
\begin{equation}
\textbf{u}_{\mathrm{rms}}=\sqrt{\frac{\left\langle\boldsymbol{u}'\cdot\boldsymbol{u}'
\right\rangle_{\Omega}}{3}},
\end{equation}
where \(\boldsymbol{\textbf{u}}'\) denotes the velocity fluctuation. And \(\varepsilon\) is the turbulent kinetic energy dissipation rate, which is calculated as:
\begin{equation}
\varepsilon= 2\nu \left\langle\sum_{i=1}^{3}\sum_{j=1}^{3} S_{ij}^{2}\right\rangle_{\Omega}.
\end{equation}
The components of the strain--rate tensor, $S_{ij}$, are defined as:
\begin{equation}
S_{ij}= \frac{1}{2}\left(\frac{\partial u_i}{\partial x_j}+\frac{\partial u_j}{\partial x_i}\right),
\qquad i,j=1,2,3.
\end{equation}
Here, $u_i$ denotes the $i$-th velocity component and $x_i$ denotes the corresponding spatial coordinate. 

\begin{figure}[H]
    \centering
    \includegraphics[width=0.6\textwidth]{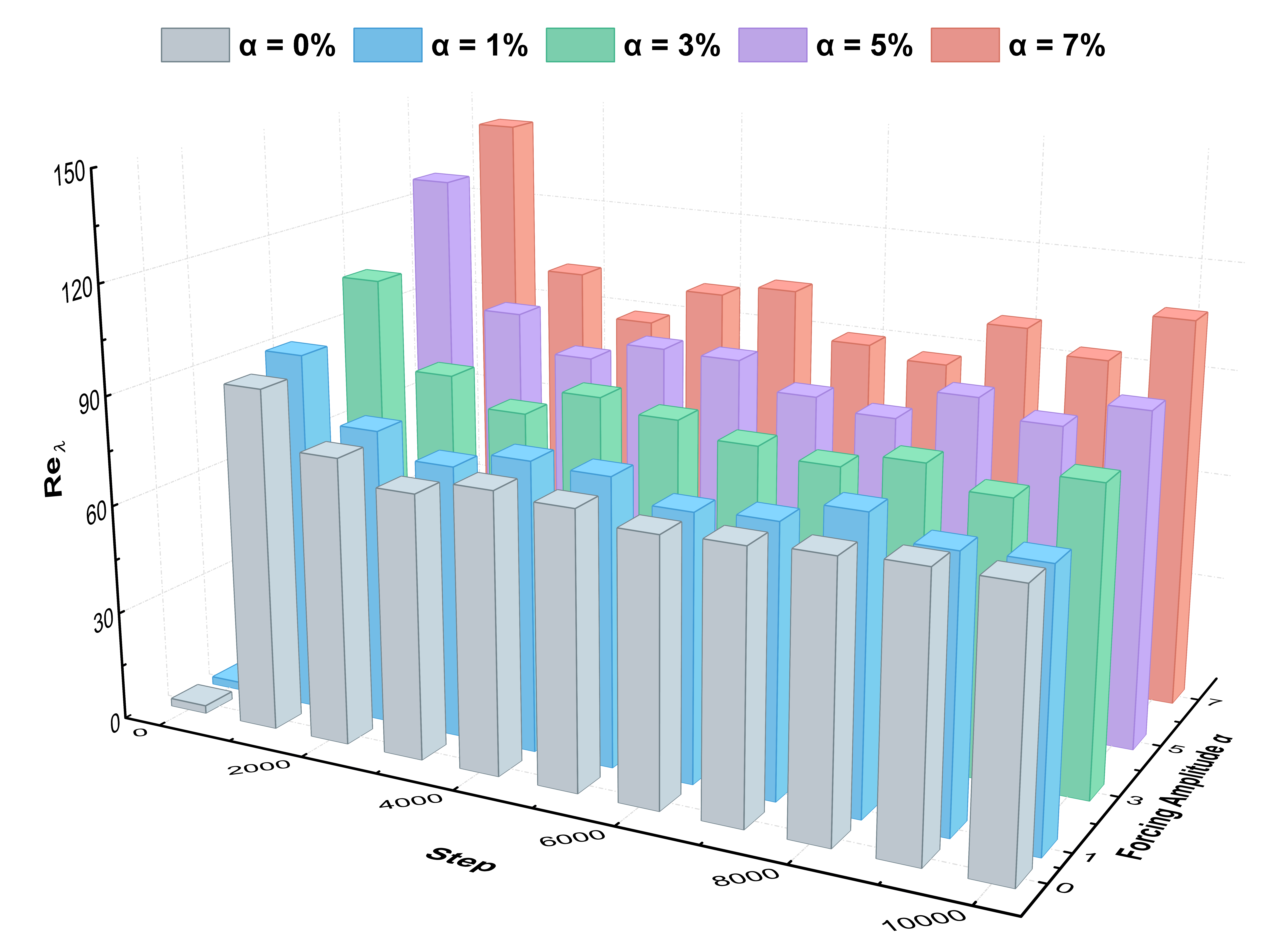}
\caption{Temporal evolution of the Taylor-microscale Reynolds number \(Re_\lambda\) for different intermediate-scale feedback coefficients \(\alpha\).}
    \label{fig:Re_lam}
\end{figure}

Figure~\ref{fig:Re_lam} shows the time evolution of \(Re_\lambda\) at different feedback coefficients. In all cases, \(Re_\lambda\) rises rapidly during the initial stage, reaches a peak, and then decreases to fluctuate around a relatively stable level. The peak value increases monotonically with feedback strength, reaching approximately \(93.95\), \(98.84\), \(111.79\), \(133.57\), and \(144.51\) for \(\alpha=0\%\), \(1\%\), \(3\%\), \(5\%\), and \(7\%\), respectively. 

After the initial transient subsides, \(Re_\lambda\) exhibits relatively stable fluctuations in all cases, and the curves retain a clear ordering with feedback strength. During the statistically stationary interval, the mean values of \(Re_\lambda\) are approximately \(73.94\), \(75.18\), \(80.54\), \(87.68\), and \(97.86\) for \(\alpha=0\%\), \(1\%\), \(3\%\), \(5\%\), and \(7\%\), respectively. Compared to the baseline without intermediate-scale feedback, the mean values in the four non-zero feedback cases increase by approximately \(1.68\%\), \(8.93\%\), \(18.59\%\), and \(32.36\%\). These results show that the feedback-induced increase in \(Re_\lambda\) persists beyond the initial transient, with its statistical mean increasing monotonically with the feedback strength.
This is because intermediate-scale feedback is applied along the local direction of the filtered velocity. Increasing the feedback coefficient \(\alpha\) raises the upper bound on the magnitude of the local feedback force, helping to strengthen velocity fluctuations at the corresponding scales. These motions exchange energy across scales through nonlinear interactions, allowing the feedback effects to extend to the overall velocity field. At fixed viscosity, \(Re_\lambda\) depends on the product of the root-mean-square velocity fluctuation and the Taylor microscale. The combined effects of feedback on velocity amplitude and spatial structure therefore explain the increase in \(Re_\lambda\).


To further examine whether increasing the feedback strength is accompanied by enhanced small scale vortical activity, we analyze the temporal evolution of enstrophy, which is defined as:
\begin{equation}
\Omega(t)=\frac{1}{2}\left\langle\boldsymbol{\omega}\cdot\boldsymbol{\omega}
\right\rangle_{\Omega},
\end{equation}
with the vorticity given by
\begin{equation}
\boldsymbol{\omega}=\nabla\times\mathbf{u}.
\end{equation} 
Because enstrophy is directly related to squared velocity gradients it is more sensitive to high-wavenumber contributions.

\begin{figure}[H]
    \centering
    \includegraphics[width=0.6\textwidth]{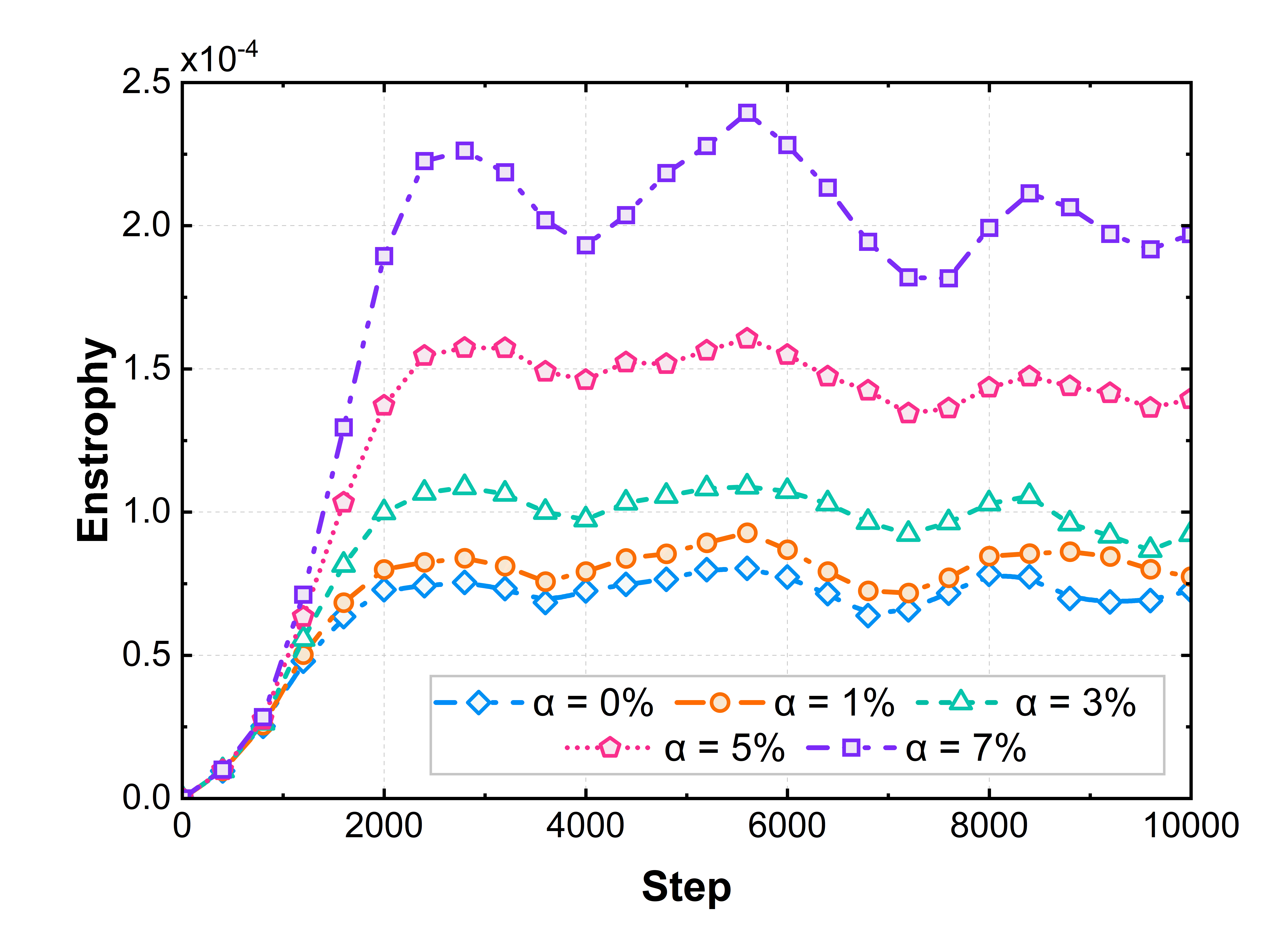}
\caption{Temporal evolution of the enstrophy for different intermediate-scale feedback coefficients \(\alpha\).}
    \label{fig:Enstrophy}
\end{figure}
As shown in Fig.~\ref{fig:Enstrophy}, enstrophy rises rapidly during the initial stage in all cases and subsequently exhibits relatively stable fluctuations. The curves remain clearly ordered by feedback coefficient \(\alpha\), with stronger-feedback cases generally exhibiting higher enstrophy levels. During the statistically stationary interval, the mean enstrophy values are approximately \(7.28\times10^{-5}\), \(8.21\times10^{-5}\), \(9.91\times10^{-5}\), \(1.45\times10^{-4}\), and \(2.05\times10^{-4}\) for \(\alpha=0\%\), \(1\%\), \(3\%\), \(5\%\) and \(7\%\), respectively. Compared to the baseline without intermediate-scale feedback, the mean values in the four non-zero feedback cases increase by approximately \(12.80\%\), \(36.04\%\), \(99.09\%\), and \(181.98\%\). These results show that the feedback-induced enhancement of vorticity activity persists throughout the statistically interval, with the mean enstrophy increasing monotonically with feedback coefficient \(\alpha\).
The increase in enstrophy is consistent with the previously observed rise in high-wavenumber energy. Since vorticity involves spatial derivatives of velocity, enstrophy is particularly sensitive to high-wavenumber velocity components. The higher energy at these wavenumbers is therefore also reflected in the vorticity statistics. In particular, as \(\alpha\) increases from \(5\%\) to \(7\%\), the additional target-band energy gain is limited, whereas the high-wavenumber energy and enstrophy continue to increase markedly. This indicates that the response at smaller scales continues to strengthen despite the limited additional gain within the target band. Meanwhile, the previously reported net forward nonlinear energy flux across \(K=12\) also continues to increase, indicating that these changes are accompanied by stronger energy transfer toward smaller scales.

\begin{figure}[H]
    \centering
    \includegraphics[width=0.6\textwidth]{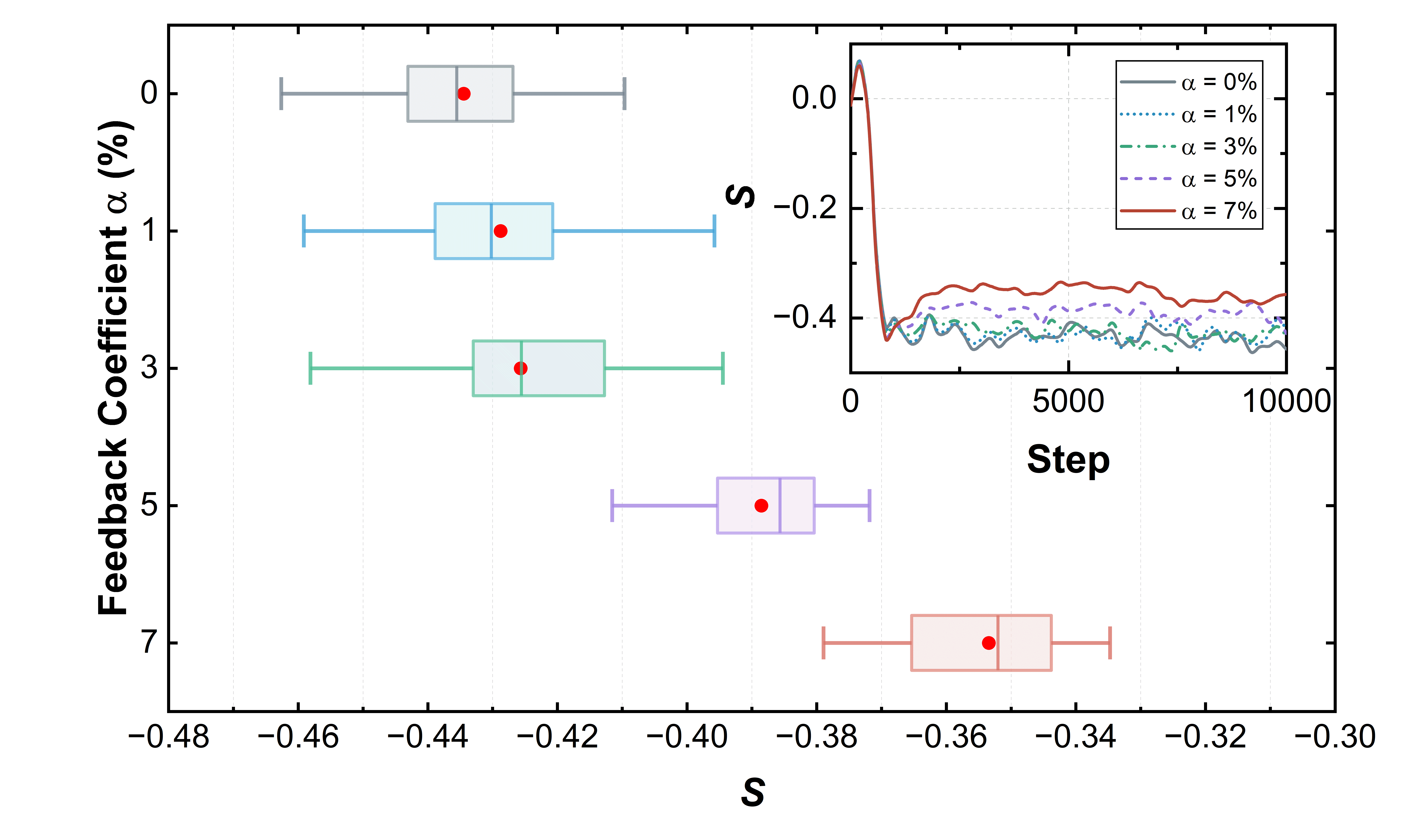}
\caption{Distribution of the longitudinal velocity-gradient skewness \(S\) for different intermediate-scale feedback coefficients \(\alpha\). The box plots summarize the statistical distribution of \(S\) over the sampled interval, with the red markers indicating the mean values. The inset shows the corresponding temporal evolution of \(Skewness\).}
    \label{fig:Skewness}
\end{figure}

To further examine the influence of feedback strength on local velocity gradient statistics, we analyze the skewness \(S\) of the longitudinal velocity gradient, defined as:
\begin{equation}
S=\frac{\left\langle\left(\partial u/\partial x\right)^3\right\rangle}{
\left\langle\left(\partial u/\partial x\right)^2\right\rangle^{3/2}}.
\end{equation}
Negative skewness indicates an asymmetry in the longitudinal velocity-gradient distribution, with a higher absolute value corresponding to a more pronounced negative asymmetry. As shown in Fig.~\ref{fig:Skewness}, following a rapid initial adjustment, the skewness remains negative and fluctuates within a relatively stable range in all feedback cases. As the feedback coefficient \(\alpha\) increases, the mean skewness shifts to zero, indicating a reduced negative asymmetry.

During the statistically stationary interval, the mean skewness values are approximately \(-0.434\), \(-0.429\), \(-0.426\), \(-0.389\), and \(-0.353\) for \(\alpha=0\%\), \(1\%\), \(3\%\), \(5\%\), and \(7\%\), respectively. The mean values for the \(0\%\), \(1\%\), and \(3\%\) cases are relatively close, with obvious overlap between their boxes in the plot. This indicates small changes in skewness under weak to moderate feedback. In contrast, the boxes and mean values for the \(5\%\) and \(7\%\) cases shift further toward zero, indicating a more pronounced effect of stronger feedback on the asymmetry of the longitudinal velocity-gradient distribution. 
These results indicate that stronger feedback coefficient \(\alpha\) reduces the imbalance between the contributions of positive and negative longitudinal velocity gradients. However, negative contributions remain dominant, so the negative asymmetry persists. Based on the mean values, the skewness magnitude at \(\alpha=7\%\) is approximately \(18.66\%\) lower than that of the baseline, indicating a more evident change in statistical asymmetry under stronger feedback. At weak to moderate feedback strengths, the mean skewness values are close and the boxes overlap. Under stronger feedback, both the mean values and boxes shift toward zero, indicating that the effect on this higher-order statistic is more evident at higher feedback strengths.

\begin{figure}[H]
    \centering
    \includegraphics[width=0.6\textwidth]{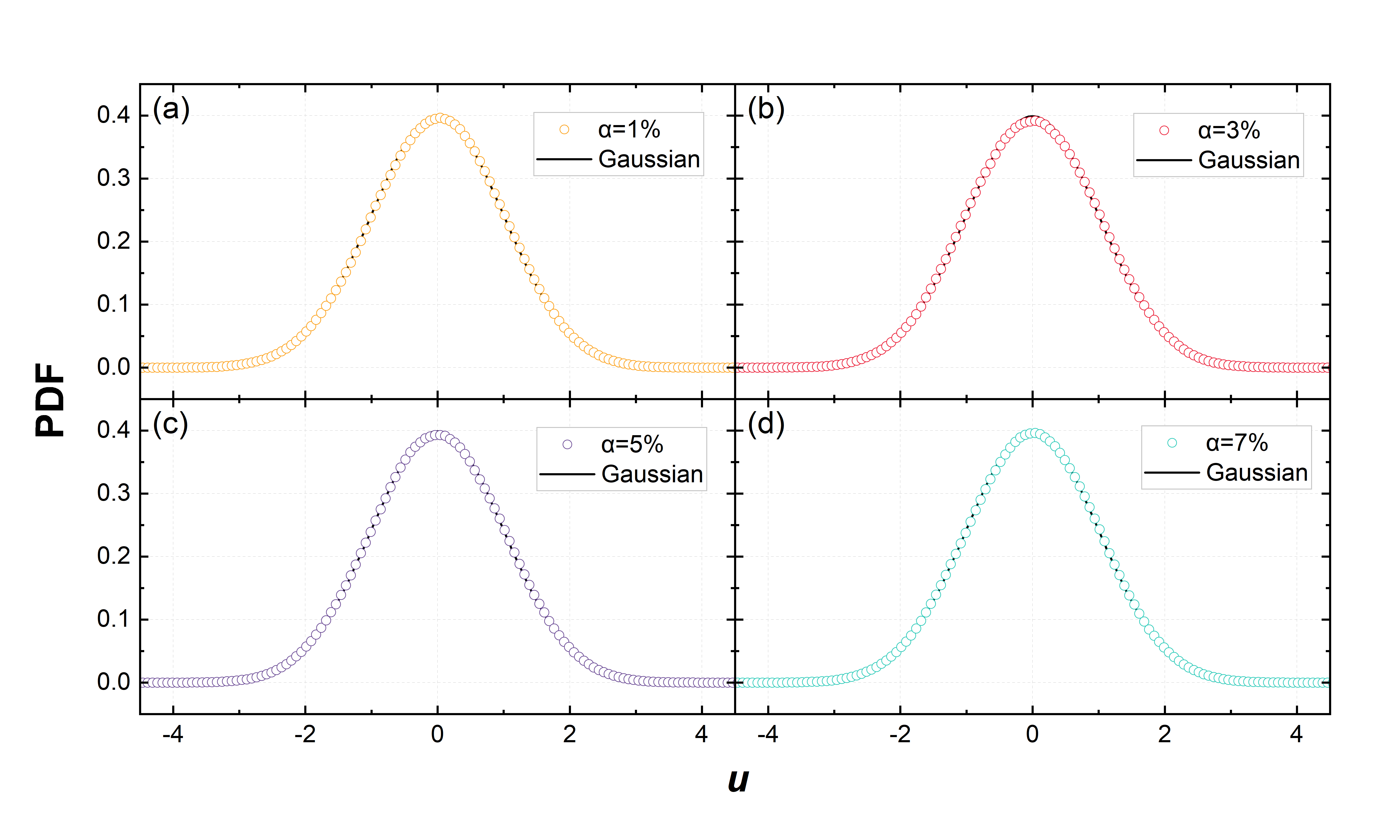}
\caption{Probability density functions of the velocity component \(u\) for different intermediate-scale feedback coefficients: (a) \(\alpha=1\%\), (b) \(\alpha=3\%\), (c) \(\alpha=5\%\), and (d) \(\alpha=7\%\). The solid black lines denote the corresponding Gaussian reference distributions.}
    \label{fig:PDF}
\end{figure}

Fig.~\ref{fig:PDF} compares the standardized single-point velocity PDFs at different feedback coefficients with the standard Gaussian distribution. Over the range shown, all standardized velocity PDFs closely follow the Gaussian reference curve. Even in the strongest feedback, \(\alpha=7\%\), the distribution remains approximately Gaussian. These results indicate that the standardized single-point velocity distributions retain a similar overall shape across the feedback strengths considered.

\begin{figure}[H]
    \centering
    \includegraphics[width=0.6\textwidth]{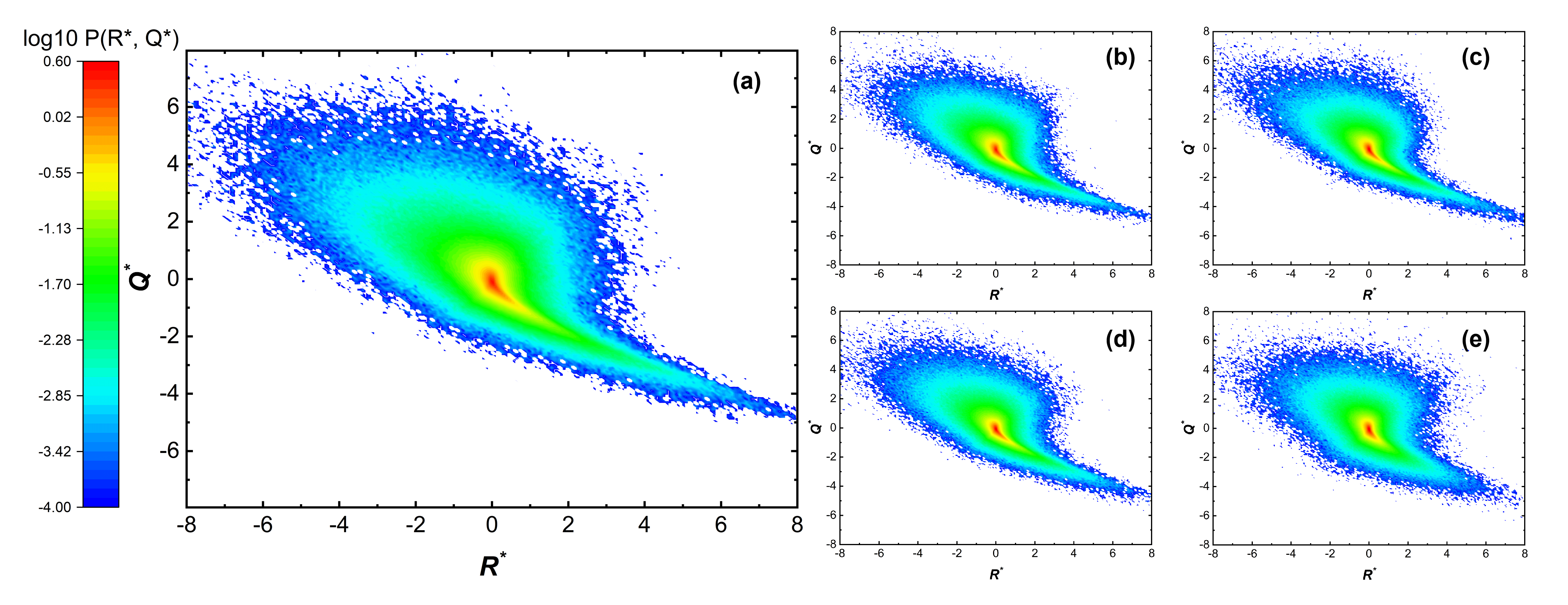}
\caption{Joint probability density distributions of the normalized velocity-gradient invariants \(Q'\) and \(R'\) for different intermediate-scale feedback coefficients: (a) \(\alpha=0\%\), (b) \(\alpha=1\%\), (c) \(\alpha=3\%\), (d) \(\alpha=5\%\), and (e) \(\alpha=7\%\). The color scale represents \(\log_{10} P(R',Q')\).}
    \label{fig:RQ}
\end{figure}

Changes in local velocity-gradient topology under intermediate-scale feedback are characterized using the joint probability density function of the normalized invariants \(Q'\) and \(R'\). As shown in Fig.~\ref{fig:RQ}, all cases retain an overall teardrop-shaped distribution, with a discernible main body and a tail extending toward positive \(R'\) and negative \(Q'\). This indicates that the basic shape of the normalized joint distribution is preserved across the feedback coefficients considered.
However, similar overall shapes do not imply identical joint statistics of velocity-gradient topology. Compared with the other panels, the main high-probability region in Fig.~\ref{fig:RQ}(e) appears visually fuller, and the regions enclosed by some contours in normalized \(Q\)--\(R\) space are somewhat broader. This suggests some redistribution of probability among velocity-gradient topological states under stronger feedback. Thus, although the principal structure of the joint distribution is largely preserved, stronger feedback alters the relative statistical weights of local flow events and broadens the range of fluctuations in the normalized velocity-gradient invariants.

\begin{figure}[H]
    \centering
    \includegraphics[width=0.6\textwidth]{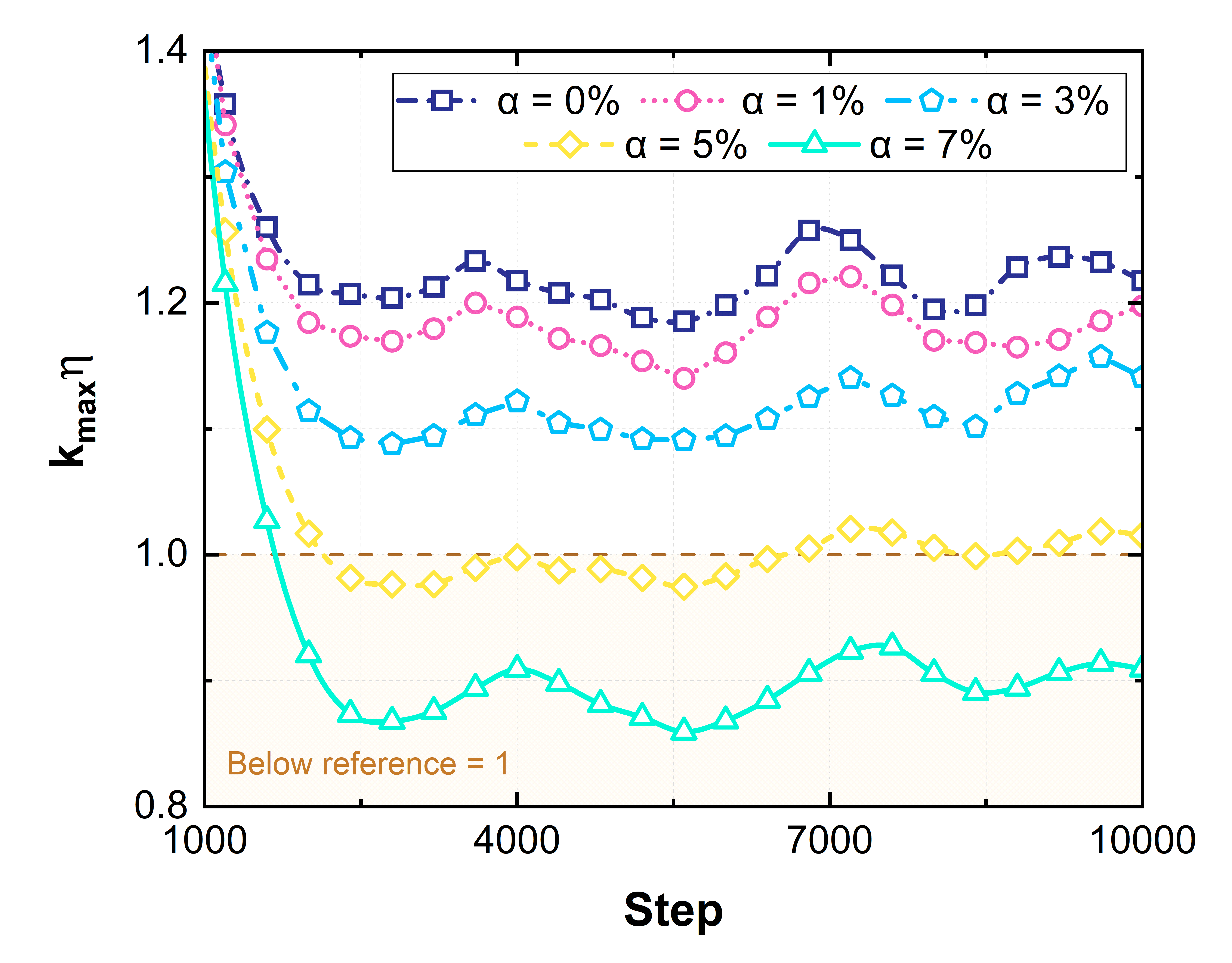}
\caption{Temporal evolution of the small-scale resolution indicator \(k_{\max}\eta\) for different intermediate-scale feedback coefficients \(\alpha\). Here, \(k_{\max}\) denotes the maximum resolved wavenumber, and \(\eta\) is the Kolmogorov length scale. The dashed horizontal line marks \(k_{\max}\eta=1\), and the shaded region highlights values below this reference level.}
    \label{fig:kmax_eta_eff}
\end{figure}

As shown in Fig.~\ref{fig:kmax_eta_eff}, \(k_{\max}\eta\) decreases systematically with increasing feedback strength. Over the statistically stationary interval, its mean values are approximately \(1.21\), \(1.18\), \(1.12\), \(1.003\), and \(0.89\) for \(\alpha=0\%\), \(1\%\), \(3\%\), \(5\%\), and \(7\%\), respectively. Relative to the no-feedback baseline, the mean values in the four nonzero-feedback cases decrease by approximately \(3.12\%\), \(8.05\%\), \(17.65\%\), and \(26.27\%\), respectively. With the reference viscosity and grid resolution held fixed across cases, this systematic decrease indicates that stronger feedback enhances small-scale activity and high-wavenumber dissipation, reducing the Kolmogorov length scale and placing increasingly stringent resolution demands on the computational grid.

The \(\alpha=0\%\), \(1\%\), and \(3\%\) cases maintain \(k_{\max}\eta>1\) over the statistically stationary interval. Their typical fluctuation ranges are approximately \(1.19\text{--}1.26\), \(1.14\text{--}1.22\), and \(1.09\text{--}1.16\), respectively. These values indicate that the current \(128^3\) grid retains some resolution margin for the dissipative scales under weak to moderate feedback. In contrast, \(k_{\max}\eta\) fluctuates around unity in the \(\alpha=5\%\) case, ranging approximately from \(0.97\) to \(1.02\), indicating proximity to the resolution limit of the present grid. At \(\alpha=7\%\), \(k_{\max}\eta\) remains below unity throughout the statistically stationary interval, with typical values of approximately \(0.86\text{--}0.93\) and a mean of about \(0.90\). This suggests that the smallest dissipative scales under the strongest feedback are no longer fully resolved by the current grid. Overall, the resolution remains adequate for \(\alpha\leq3\%\), while the \(\alpha=5\%\) case is marginally resolved and the \(\alpha=7\%\) case exhibits clear under-resolution. Under the present computational conditions, the \(\alpha=5\%\) case therefore retains a larger resolution margin than the \(7\%\) case and can serve as a representative case balancing the target-band response and resolution requirements.

In summary, intermediate-scale feedback altered both the overall turbulence state and certain local flow statistics. As \(\alpha\) increased, the Taylor microscale Reynolds number \(Re_\lambda\) and enstrophy both increased, reflecting changes in the overall turbulence state and enhanced vorticity activity. Meanwhile, longitudinal velocity-gradient skewness shifted toward zero, indicating reduced negative asymmetry. The standardized velocity PDFs remained close to the Gaussian reference distribution, and the normalized \(Q'\)--\(R'\) joint distributions retained their overall teardrop shape. However, the main body of the joint distribution appeared somewhat broader in the strongest-feedback case.

Stronger feedback also imposed increasingly stringent small-scale spatial resolution requirements. The \(\alpha=5\%\) case approached the resolution limit of the current grid, whereas the \(\alpha=7\%\) case was under-resolved and yielded only a limited additional gain in target-band energy. We therefore select \(\alpha=5\%\) as a representative compromise between target-band response and resolution constraints under the present computational conditions. In the following section, we compare this case with an enhanced large-scale forcing control at approximately matched mean dissipation rates.

\subsection{Comparison of forcing strategies under approximately matched power input}

Adding intermediate-scale feedback could also affect the total energy input. Comparing the composite forcing case only with the baseline without intermediate-scale feedback is not enough to rule out the effect of increased total input power on these responses. An additional comparison is therefore needed to test whether the observed enhancements result solely from increased total input power. For this purpose, a case with large-scale forcing alone is constructed. The forcing amplitude is adjusted so that the mean dissipation rate $\overline{\varepsilon}$ closely matches that of the  intermediate-scale feedback case at $\alpha=5\%$.

\begin{figure}[H]
    \centering
    \includegraphics[width=0.6\textwidth]{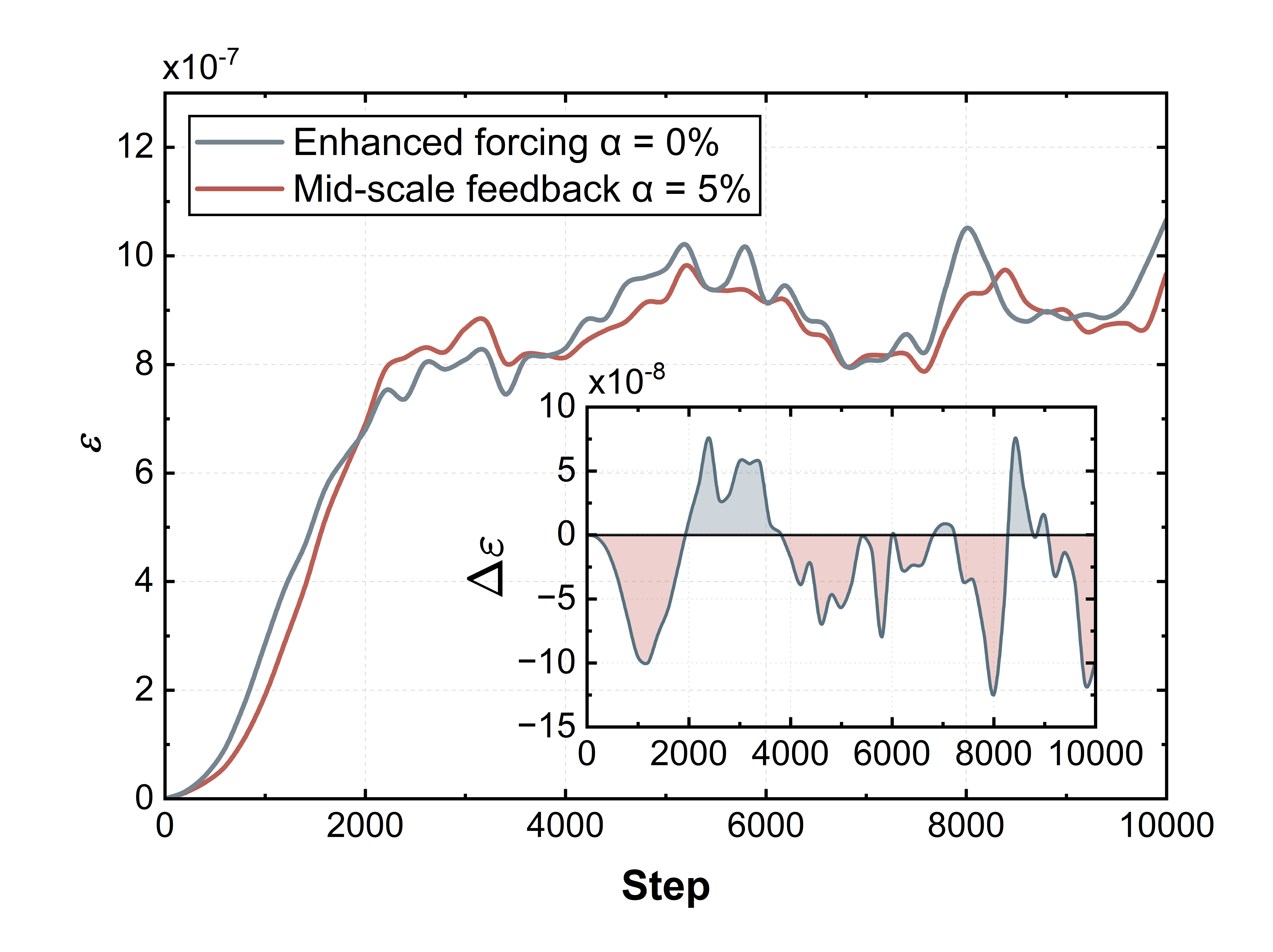}
\caption{Temporal evolution of the dissipation rate \(\varepsilon\) for the enhanced large-scale-forcing control case and the \(\alpha=5\%\) intermediate-scale-feedback case. The inset shows the instantaneous dissipation difference \(\Delta\varepsilon\) between the two cases.}
    \label{fig:Disspation}
\end{figure}

If the mean energy budget is approximately balanced under statistically stationary conditions, the mean dissipation rate \(\overline{\varepsilon}\) can be considered as an measure of the total input power. Fig.~\ref{fig:Disspation} compares the temporal evolution of the dissipation rates \(\varepsilon\) in the two cases. Following the initial development stage, the curves show similar overall levels during the statistically stationary interval. Their mean dissipation rates \(\overline{\varepsilon}\) are \(9.19\times10^{-7}\) and \(8.90\times10^{-7}\) for the enhanced large-scale forcing and intermediate-scale feedback cases, respectively, corresponding to a difference of approximately \(3.26\%\) relative to the latter. The inset shows the instantaneous dissipation difference \(\Delta\varepsilon= \varepsilon_{\mathrm{mid}}(t)
- \varepsilon_{\mathrm{enh}}(t)\), which fluctuates and changes sign repeatedly, while its time-averaged offset is small relative to the mean dissipation level. Thus, although the instantaneous dissipation rates \(\varepsilon\) are not identical, their mean levels are approximately matched, providing a reference for the scale-wise responses of the two forcing strategies under approximately equal mean input power.

\begin{figure}[H]
    \centering
    \includegraphics[width=0.6\textwidth]{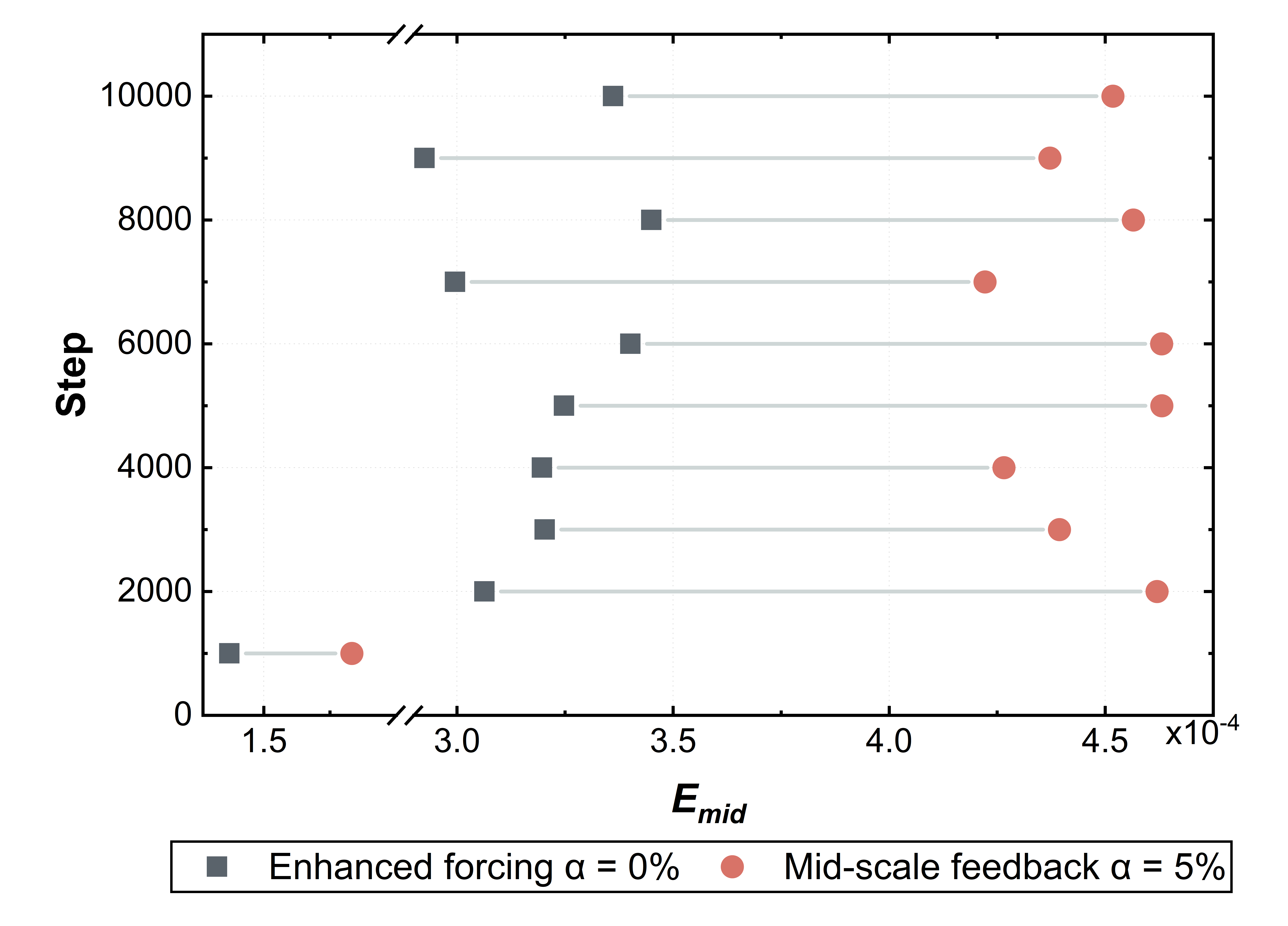}
\caption{Comparison of the target intermediate-scale energy \(E_{\rm mid}\) between the enhanced large-scale-forcing control case and the \(\alpha=5\%\) intermediate-scale-feedback case at identical sampling instants. The horizontal connecting lines link the two cases at the same time step.}
    \label{fig:Emid}
\end{figure}

The effects of the two forcing strategies on intermediate-scale energy are further compared. Fig.~\ref{fig:Emid} shows \(E_{\rm mid}\) at the same sampling times for the enhanced large-scale forcing case and the intermediate-scale feedback case at \(\alpha=5\%\). During the statistically stationary interval, \(E_{\rm mid}\) is higher in the intermediate-scale feedback case. The mean intermediate-scale energies of the enhanced large-scale forcing case and the $\alpha=5\%$ intermediate-scale feedback case are approximately $3.33\times10^{-4}$ and $4.39\times10^{-4}$, respectively. The mean intermediate-scale energy is approximately $31.83\%$ higher in the feedback case. Despite its slightly lower mean dissipation rate \(\overline{\varepsilon}\), the intermediate-scale feedback case maintains a higher mean target-band energy. This indicates that intermediate-scale feedback produces a stronger intermediate-scale energy response than enhancing large-scale forcing alone.

\begin{figure}[H]
    \centering
    \includegraphics[width=0.6\textwidth]{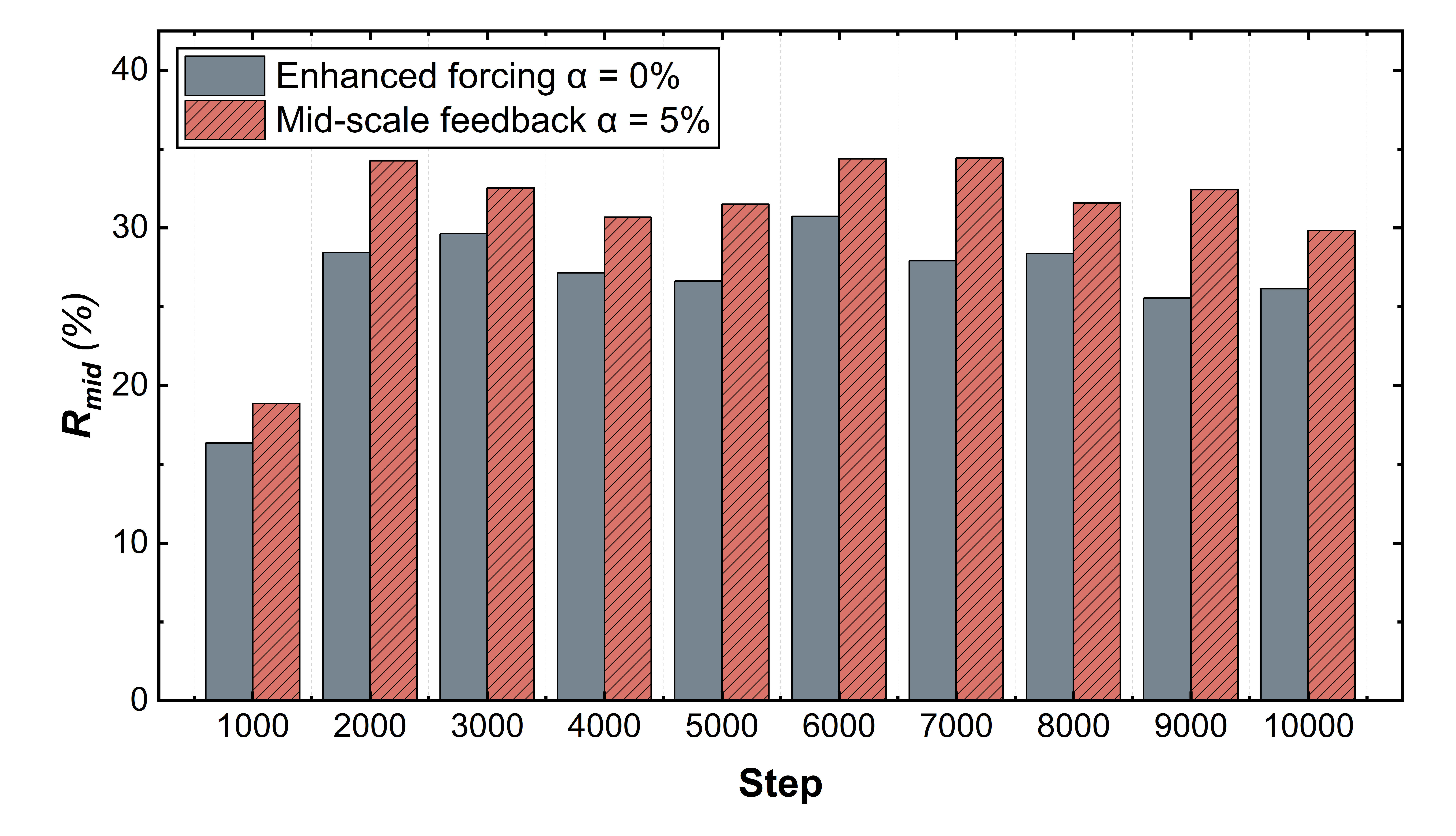}
\caption{Temporal evolution of the target intermediate-scale energy fraction \(R_{\rm mid}\) for the enhanced large-scale-forcing control case and the \(\alpha=5\%\) intermediate-scale-feedback case. \(R_{\rm mid}\) denotes the fraction of the total kinetic energy contained within the target intermediate-scale band.}
    \label{fig:Bandlast}
\end{figure}

To further characterize the effects of the two forcing strategies on the energy distribution across scales, Fig.~\ref{fig:Bandlast} compares the target intermediate-scale energy fraction \(R_{\rm mid}\) in the enhanced large-scale forcing case and the intermediate-scale feedback case at \(\alpha=5\%\). The intermediate-scale feedback case has a higher \(R_{\rm mid}\) at each sample time. During the sampling interval, the mean fractions are approximately \(27.55\%\) and \(32.36\%\), respectively, corresponding to a difference of \(4.81\) percentage points and a relative increase of approximately \(17.45\%\) in the latter case. The relative increases from approximately \(11.39\%\) to \(26.92\%\), indicating that the difference persists throughout the sample times. 

\begin{figure}[H]
    \centering
    \includegraphics[width=0.6\textwidth]{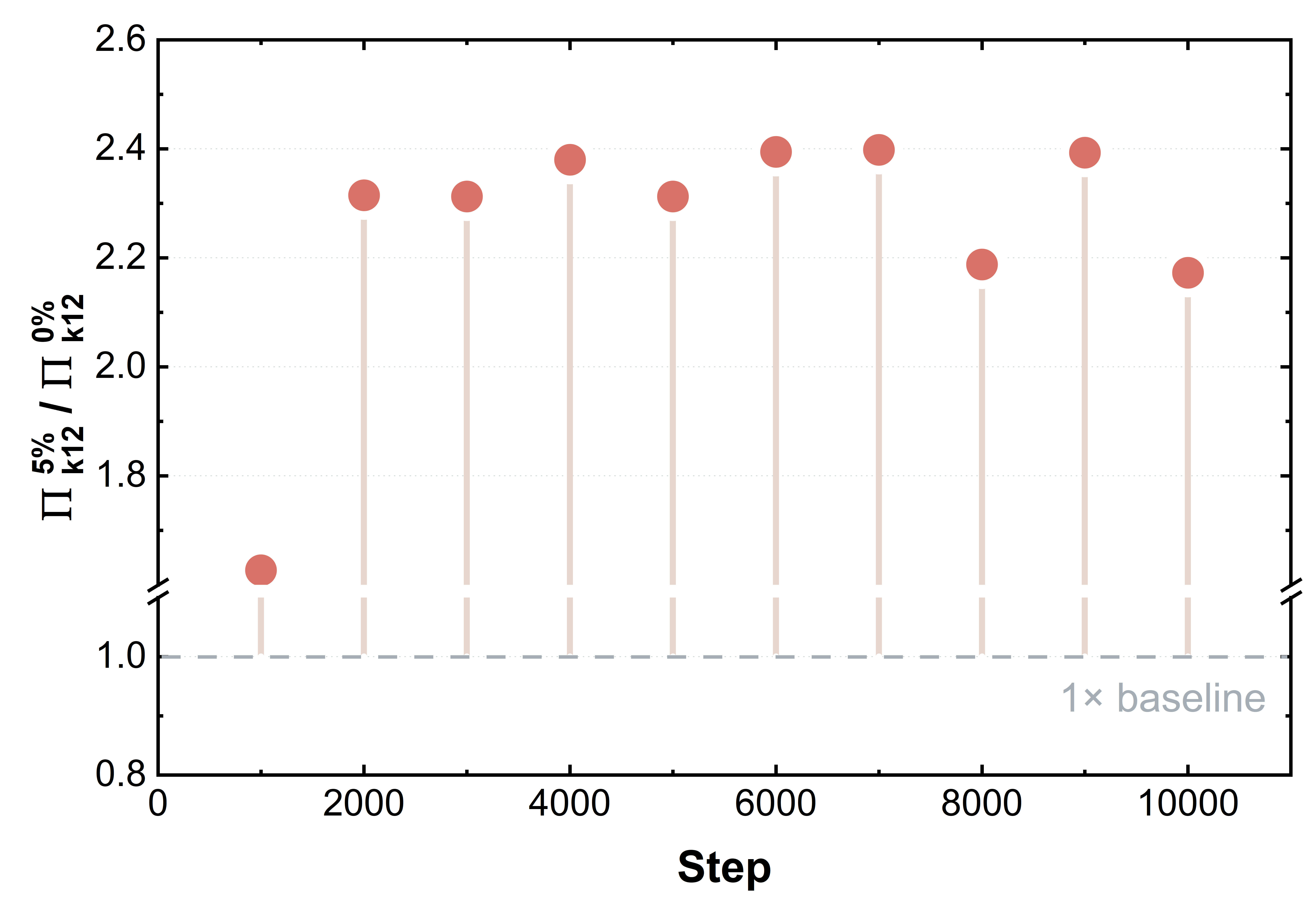}
\caption{Temporal evolution of the ratio \(\Pi_{12}^{5\%}/\Pi_{12}^{0\%}\) between the cumulative nonlinear energy flux across \(K=12\) in the \(\alpha=5\%\) intermediate-scale-feedback case and that in the enhanced large-scale-forcing control case. The dashed line at unity denotes equal flux levels between the two cases.}
    \label{fig:Nonlinear}
\end{figure}

To examine whether the enhancement of nonlinear energy transfer persists under approximately matched mean dissipation rates \(\overline{\varepsilon}\), Fig.~\ref{fig:Nonlinear} compares the cumulative nonlinear energy flux at \(K=12\) between the intermediate-scale feedback case at \(\alpha=5\%\) and the enhanced large-scale forcing control case. It also shows their ratio, \(\Pi_{12}^{5\%}/\Pi_{12}^{ 0\%} \), where the denominator denotes the flux in the enhanced large-scale forcing control case. The ratio is clearly greater than \(1\) at all times, indicating that the intermediate-scale feedback case has a higher net forward nonlinear energy flux across the high-wavenumber boundary of the target band. Beyond the early stage of development, the ratio generally remains within approximately \(2.17\text{--}2.39\). This difference indicates that comparable global dissipation levels do not imply identical nonlinear energy transfer. The scales at which the forcing acts still influence energy transport toward smaller scales. Compared with enhancing large-scale forcing alone, intermediate-scale feedback more effectively promotes net forward nonlinear energy transfer across $K=12$. This supports the strategy's ability to selectively enhance energy transfer across the high-wavenumber boundary of the target band.

In summary, the comparison shows that the scales at which forcing stages can influence energy distribution and transfer. At approximately matched mean dissipation rates, the intermediate-scale feedback case $\alpha=5\%$ has a higher mean target-band energy and a larger target-band energy fraction than the enhanced large-scale forcing case. The increases in both quantities indicate a stronger target-band energy response and a greater relative contribution of this band to the total energy. Despite its slightly lower mean dissipation rate, the intermediate-scale feedback case maintains a higher target-band energy, indicating a more effective intermediate-scale control. The net forward nonlinear energy flux across $K=12$ is also higher, and this difference persists throughout the samples. These results indicate that an increase in target-band energy occurs along with enhanced net nonlinear energy transfer toward smaller scales.

\section{Conclusions}
\label{sec:conclusions}

This work proposes a composite large-scale and intermediate-scale forcing strategy for the LBM simulation of the FHIT. The stage keeps traditional low-wavenumber forcing to maintain turbulence and uses composite filtering in physical space to extract the velocity field to construct intermediate-scale feedback. The feedback force is applied along the local direction of the filtered velocity, with its magnitude bounded by a prescribed fraction of the local large-scale forcing magnitude. The selected filter configuration \(80+20\) provides a reasonable compromise between the concentration of energy in the target band and the retention of the target band response. For the nonzero feedback strengths examined, the spatially averaged construction power \(P_{\mathrm{mid}}\) between the feedback force and the filtered velocity remains positive during the statistically stationary sampling interval. The time-averaged fraction of the cross-power spectral magnitude within the target band \(5\leq k\leq12\) ranges from \(72.26\%\) to \(74.56\%\).

The energy spectra and band-integrated energies show that intermediate-scale feedback increases energy in both the prescribed target band and the high-wavenumber band, with a larger relative increase in the latter. Compared to the baseline without intermediate-scale feedback, the mean target-band energy increases by approximately $6.29\%$, $11.65\%$, $19.89\%$, and $18.99\%$ for $\alpha=1\%$, $3\%$, $5\%$ and $7\%$, respectively. Nonlinear energy flux analysis further shows that these changes in energy distribution are accompanied by enhanced net forward nonlinear energy transfer. The relative flux enhancement across \(K=12\) is greater than that across \(K=4\), indicating a scale-dependent transfer response. For $\alpha=1\%$, $3\%$, $5\%$, and $7\%$, the mean nonlinear energy flux across $K=12$ reaches $1.18$, $1.54$, $2.32$, and $3.06$ times that of the original baseline without intermediate-scale feedback, respectively. These results indicate that net forward nonlinear energy transfer increases with feedback strength.

The standardized velocity PDFs remain close to the Gaussian reference distribution in all cases, and the normalized \(Q'\)--\(R'\) joint distributions retain their overall teardrop shape, although the main region exhibits some broadening under the strongest feedback. Meanwhile, the mean skewness of the longitudinal velocity gradient shifts toward zero as feedback increases, indicating reduced negative statistical asymmetry. As \(\alpha\) increases from \(1\%\) to \(7\%\), the Taylor-microscale Reynolds number \(Re_\lambda\) and enstrophy continue to increase. The mean $Re_\lambda$ is $73.94$ in the baseline case without intermediate-scale feedback. For $\alpha=1\%$, $3\%$, $5\%$, and $7\%$, the corresponding mean values are $75.18$, $80.54$, $87.68$, and $97.86$, respectively. Compared to baseline without intermediate-scale feedback, enstrophy increases by $12.80\%$, $36.04\%$, $99.09\%$, and $181.98\%$ for $\alpha=1\%$, $3\%$, $5\%$, and $7\%$, respectively. This increasing trend reflects the continued response of the overall turbulence state as the feedback strength increases. However, the small-scale resolution indicator \(k_{\max}\eta\) generally decreases as the feedback coefficient \(\alpha\) increases. During the plotted sampling interval, the minimum \(k_{\max}\eta\) values are \(1.18\), \(1.14\), \(1.09\), \(0.97\), and \(0.86\) for \(\alpha=0\%\), \(1\%\), \(3\%\), \(5\%\), and \(7\%\), respectively. The minimum values in the first three cases exceed the reference threshold of unity, indicating that some resolution margin remains. In the \(\alpha=5\%\) case, \(k_{\max}\eta\) fluctuates around the reference line and falls below the threshold in the samples. This indicates that the case is close to the resolution limit of the current grid. In contrast, \(k_{\max}\eta\) remains below unity at late times in the \(\alpha=7\%\) case. According to the criterion adopted here, this case is under-resolved at small scales, and its response should be interpreted in light of this resolution limitation. Under the present computational conditions, this case thus serves as a representative compromise between the target-band response and resolution constraints.

Comparison with the enhanced large-scale forcing case further shows that the two forcing strategies produce different responses on a scales. Although the mean dissipation rates \(\varepsilon\) of the two cases differ by only approximately \(3.26\%\), the mean target-band energy in the intermediate-scale feedback case \(\alpha=5\%\) is approximately \(31.83\%\). Its mean target-band energy fraction is \(32.36\%\), compared to \(27.55\%\) in the enhanced large-scale forcing case. At samples after the initial transient, the net forward nonlinear energy flux across \(K=12\) in the intermediate-scale feedback case generally remains approximately \(2.17\)--\(2.39\) times.

In summary, the proposed combined large- and intermediate-scale forcing strategy retains large-scale forcing to sustain turbulence and constructs scale-selective feedback through composite physical-space filtering. It provides a way to regulate target-band energy distribution and net forward nonlinear energy transfer toward smaller scales. Under approximately matched mean dissipation rates, the strategy maintains a higher target-band energy and a larger target-band energy fraction than enhanced large-scale forcing alone. It also more effectively promotes net forward nonlinear energy transfer across the high-wavenumber boundary of the target band. These results highlight the role of forcing scale selectivity in energy control. Stronger feedback also changes the overall turbulence state and some local flow statistics, while placing greater demands on small-scale resolution. Consequently, \(\alpha=5\%\) is selected as the preferred case for the grid, flow parameters, and target band examined here. Future studies could combine grid refinement and convergence tests to examine feedback responses at different Reynolds numbers and for different target bands. They could also explore the adaptive adjustment of the feedback strength based on target-band energy gains and resolution constraints.

\section{Acknowledgments}
This work was primarily supported by the National Natural 
Science Foundation of China (Grant No. 11872382) and the 
Thousand Young Talents Program of China.
The author sincerely acknowledges the guidance and generous 
support of Professor Ewe-Wei Saw whose insightful discussions and continuous encouragement were invaluable to the completion of this work.
Special thanks are extended to Ms. Li Zhang for her 
dedicated administrative and logistical support.
\bibliographystyle{unsrt}
\bibliography{References}

\end{document}